\documentclass[prl,aps,reprint,superscriptaddress,twocolumn,letter,amssymb,amsfonts,amsmath,nopacs,nofootinbib,longbibliography]{revtex4-2}
\usepackage{graphicx}
\usepackage{bm}
\usepackage[dvipsnames]{xcolor}
\usepackage{comment}
\usepackage{url}
\usepackage{epstopdf}
\usepackage{amsthm,mathrsfs} 

\usepackage{txfonts}
\usepackage[colorlinks=true,linkcolor=blue,urlcolor=blue,citecolor=blue,pdfusetitle]{hyperref}
\usepackage{soul}
\usepackage{float}
\usepackage{pgfplots}
\usepackage{caption}
\usepackage{subcaption}

\newcommand{\kt}{\tilde{\kappa}}

\usepackage[framemethod=TikZ]{mdframed}
    
    \usepackage{tikz}
\usetikzlibrary{calc}

\xdefinecolor{mycolor}{RGB}{62,96,111} 
\colorlet{bancolor}{mycolor}

\mdfdefinestyle{MyFrame}{%
    linecolor=black,
    outerlinewidth=2pt,
    innertopmargin=4pt,
    innerbottommargin=4pt,
    innerrightmargin=4pt,
    innerleftmargin=4pt,
        leftmargin = 4pt,
        rightmargin = 4pt,
    backgroundcolor=gray!50!white
        }

\AtBeginDocument{\mathcode`v=\varv}

\begin{document}
\count\footins = 1000
\title{Optical solitons in curved spacetime}

\author{Felix Spengler}
\affiliation{Institut f\"{u}r Theoretische Physik, Eberhard-Karls-Universit\"{a}t T\"{u}bingen, 72076 T\"{u}bingen, Germany}
\author{Alessio Belenchia}
\affiliation{Institut f\"{u}r Theoretische Physik, Eberhard-Karls-Universit\"{a}t T\"{u}bingen, 72076 T\"{u}bingen, Germany}
\affiliation{Centre for Theoretical Atomic, Molecular, and Optical Physics, School of Mathematics and Physics, Queens University, Belfast BT7 1NN, United Kingdom}
\author{Dennis R\"{a}tzel}
 \affiliation{ZARM, University of Bremen, Am Fallturm 2, 28359 Bremen, Germany}
\affiliation{Humboldt  Universit\"{a}t  zu  Berlin,  Institut  f\"{u}r  Physik, Newtonstraße  15,  12489  Berlin,  Germany}
\author{Daniel Braun}
\affiliation{Institut f\"{u}r Theoretische Physik, Eberhard-Karls-Universit\"{a}t T\"{u}bingen, 72076 T\"{u}bingen, Germany}

\date{\today}

\begin{abstract}
Light propagation in curved spacetime is at the basis of some of the most stringent tests of Einstein's general relativity. At the same time, light propagation in media is at the basis of several communication systems. Given the ubiquity of the gravitational field, and the exquisite level of sensitivity of optical measurements, the time is ripe for investigations combining these two aspects and studying light propagation in media located in curved spacetime. 
In this work, we focus on the effect of a weak gravitational field on the propagation of optical solitons in non-linear optical media. We derive a non-linear Schr\"{o}dinger equation describing the propagation of an optical pulse in an effective, gradient-index medium in flat spacetime, encoding both the material properties and curved spacetime effects. In analyzing the special case of propagation in a 1D optical fiber, we also include the effect of mechanical deformations and show it to be the dominant effect for a fiber oriented in the radial direction in Schwarzschild spacetime.
\end{abstract}
\maketitle

\section{Introduction}
The properties of light propagating in optical media is a subject as old as optics itself. In recent years, the possibility to engineer novel metamaterials has opened the door to the so-called transformation optics~\cite{leonhardt2009transformation}, a field promising to enhance existing devices and create novel ones. At the basis of this revolution is the fact that, in the geometric optics limit -- and neglecting dispersion --, light rays propagate in media following the geodesics of an effective Lorentzian metric, the so-called optical metric~\cite{gordon1923lichtfortpflanzung}. This has also led to the investigation of light in optical media as an analogue gravity model, i.e., a model in which field perturbations propagate \textit{as if} in a curved spacetime background, particularly useful in the investigation of kinematic effects of quantum field theory in curved spacetime, like the Hawking radiation and cosmological particle production~\cite{barcelo2011analogue,philbin2008fiber,rubino2011experimental}. When also the effect of dispersion is considered, the metric description can be cast aside for a more powerful Hamiltonian formalism, giving rise to the so-called ray-optical structures~\cite{1975A&A....44..389B,perlick2000ray}. 

This analogy between optical media and curved spacetimes can be pushed even further by showing that Maxwell equations in vacuum, curved spacetime are equivalent to flat-spacetime Maxwell equations in the presence of a bi-anisotropic moving medium whose dielectric permittivity and magnetic permeability are determined entirely by the space-time metric~\cite{plebanski1960electromagnetic}. Spacetime itself can then be described as an optical medium at the level of full electromagnetism. It is then natural to wonder what would happen if light were to propagate in an optical medium placed in a curved spacetime. Far from being a far-fetched situation, this is exactly the case for light propagating in media on Earth due to the non-vanishing, albeit weak, gravitational field of our planet. In this work, we are interested in exactly this situation. In particular, while at the geometric optics level the formalism of ray-optical structures can be used, we aim here at a description, analogous to the one in~\cite{plebanski1960electromagnetic}, at the level of full Maxwell equations. Indeed, such a description allows for the modelling of the propagation of intense pulses in situations of physical interest, like soliton propagation in optical fibers, taking into account the effect of a weak gravitational field.  

We show that light propagation in a medium in curved spacetime is equivalent to propagation in an effective medium in flat spacetime. 
We then use this formalism to investigate the propagation of intense light pulses in non-linear media, giving rise to optical solitons. Solitons, and more in general propagating pulses, in optical fibers are at the basis of several communication protocols. Given that fibers on Earth are \textit{de facto} in a curved spacetime due to our planet's gravitational field, it is relevant to analyze how gravity influences light-pulses propagation. Our result allows us to set up a framework for the analysis of the effect of acceleration and curvature on the propagation of pulses in optical fibers in curved spacetimes. We numerically investigate some of these effects for the simple case of 1D propagation in the weak-field limit.

\section{An effective ``spacetime medium''}

While light in media can propagate as in a curved spacetime, curved spacetime can also be seen as an effective medium with non-trivial permeability and permittivity~\cite{plebanski1960electromagnetic,de1971gravitational}. It is not difficult to generalize the derivations in~\cite{plebanski1960electromagnetic,de1971gravitational} to the case in which light propagates in an optical medium placed in curved spacetime. Also in this case it can be shown that Maxwell's equations are equivalent to Maxwell's equations in flat spacetime for an effective medium whose properties encode both the ones of the physical medium and of curved spacetime.

Indeed, consider a dielectric and permeable medium in curved spacetime characterized by a Lorentzian metric $g_{\mu\nu}$ with mostly plus signature. We follow here the notation of~\cite{perlick2000ray}, also reported in the Supplemental Material~\cite{SM}. 
Maxwell's equations 
in the absence of free charges and currents 
are given by 
\begin{align}
    \nabla_k F^{*\,ik}=0\\
    \nabla_k G^{ik}=0,
\end{align}
where $F^*$ is the Hodge dual of the electromagnetic tensor $F$, and $G$ and $F$ are related by the constitutive equations of the material. Choosing an observer field $u^i$, the electric and magnetic field strengths can be defined with respect to it as 
\begin{align}
    & B_a=-\frac{1}{2}\eta_{abcd}u^b F^{cd};\,\,E_i=F_{ij}u^j\\
    & H_a=-\frac{1}{2}\eta_{abcd}u^b G^{cd};\,\,D_i=G_{ij}u^j\\
    &F_{ab}=-\eta^{cd}_{ab}u_d B_c+2u_{[a}E_{b]}\\
    &G_{ab}=-\eta^{cd}_{ab}u_d H_c+2u_{[a}D_{b]},
\end{align}
in the reference frame of the observer in which the medium is assumed to be at rest. Here $\eta_{ijkl}=\sqrt{-g}\delta_{ijkl}$ is the Levi-Civita tensor and $T_{[abc\dots]}$ denotes the antisymmetrization of the tensor with respect to the indices in square brackets.

As discussed in~\cite{SM}, choosing $u^{i}=\delta^i_0/\sqrt{-g_{00}}$, the projection of Maxwell's equations in 3-dimensional form leads to
\begin{align}
    &\delta^{\alpha\beta\gamma}\partial_\beta\mathcal{H}_\gamma-\partial_0\mathcal{D}^\alpha=0;\,\,\partial_l\mathcal{D}^l=0\\
    &\delta^{\alpha\beta\gamma}\partial_\beta\mathcal{E}_\gamma+\partial_0\mathcal{B}^\alpha=0;\,\,\partial_l\mathcal{B}^l=0,
\end{align}
where $\mathcal{E}_\alpha=\sqrt{-g_{00}}E_\alpha$, $\mathcal{H}_\alpha=\sqrt{-g_{00}}H_\alpha$, and
\begin{align}\label{Maxelleff}
    & \mathcal{D}^\alpha=-\sqrt{-g}\frac{g^{\alpha\beta}}{g_{00}}\mathfrak{D}_\beta-\delta^{\alpha\beta\gamma}\frac{g_{0\gamma}}{g_{00}}\mathcal{H}_\beta\\
     & \mathcal{B}^\alpha=-\sqrt{-g}\frac{g^{\alpha\beta}}{g_{00}}\mathfrak{B}_\beta+\delta^{\alpha\beta\gamma}\frac{g_{0\gamma}}{g_{00}}\mathcal{E}_\beta,
\end{align}
with $\mathfrak{B}_\alpha=\sqrt{-g_{00}}B_\alpha$, and $\mathfrak{D}_\alpha=\sqrt{-g_{00}}D_\alpha$.
These expressions are equivalent to Maxwell's equations in flat spacetime in the presence of an optical medium. In particular, for a non-dispersive medium characterized by constitutive relations $D_a=\varepsilon^b_a E_b$, and $B_a=\mu^b_a H_b$, the effective medium will be characterized by a dielectric and magnetic permeability given by the product of the material ones and the ones characterizing the curved spacetime~\cite{plebanski1960electromagnetic,de1971gravitational}. {Indeed, expressing $\mathcal{D}^\alpha=\tilde{\varepsilon}^{\alpha\beta}\mathcal{E}_\beta+\tilde{\gamma}^\beta_{\alpha}\mathcal{H}_\beta$ and correspondingly $\mathcal{B}^\alpha=\tilde{\mu}^{\alpha\beta}\mathcal{H}_\beta-\tilde{\gamma}^\beta_\alpha \mathcal{E}_\beta$, where $\tilde{\gamma}^\beta_\alpha$ encode magnetoelectric effects, we see that 
} 
\begin{align}\label{permten}
    & \tilde{\mu}^{\alpha\beta}=-\sqrt{-g}\frac{g^{\alpha\gamma}}{g_{00}}\mu_\gamma^{\;\beta}\\
    & \tilde{\varepsilon}^{\alpha\beta}=-\sqrt{-g}\frac{g^{\alpha\gamma}}{g_{00}}\varepsilon_\gamma^{\;\beta},
\end{align}
{and $\tilde{\gamma}^{\alpha\beta}=-\delta^{\alpha\beta\gamma}g_{0\gamma}/g_{00}$\footnote{Note that, in the case the material itself possesses magnetoelectric terms in the constitutive equations, i.e., $D_a=\varepsilon^b_a E_b+\gamma^b_{a}H_b$, and $B_a=\mu^b_a H_b-\gamma^b_a E_b$ then $\tilde{\gamma}^{\alpha\beta}=-\delta^{\alpha\beta\gamma}\frac{g_{0\gamma}}{g_{00}}-\sqrt{-g}\frac{g^{\alpha\delta}}{g_{00}}\gamma^{\beta}_\delta$}.}
As a direct consequence, whenever the refractive index of the \emph{effective medium} can be defined, it will also be the product of the material refractive index times the vacuum spacetime effective one. The same result can be easily obtained at the level of geometric optics.

Finally, we make two observations relevant for the study of the propagation of light pulses. Firstly, a non-magnetic material in curved spacetime corresponds to a magnetic effective medium in Minkowski due to the ``magnetic permeability'' of the background spacetime. Secondly, when considering a non-linear material, we see that the non-linearity will also be affected by the curvature of spacetime as well as the linear polarizability.

\section{Pulse propagation: Non-linear Schrödinger equation}
We next consider the propagation of light pulses in a Kerr non-linear, non-magnetic material in curved spacetime. In particular, we focus on the case in which the material is in a stationary orbit of Schwarzschild spacetime and use isotropic coordinates. This situation well-captures the cases of interest for optical communication and laboratory experiments like, e.g., optical fibers hanging still above Earth's surface.  

In flat spacetime, the non-linear Schrödinger equation (NLSE) is often used when considering the propagation of light pulses whose amplitude is well-described by a scalar envelope slowly varying with respect to the light period and wavelength~\cite{agrawal2000nonlinear,boyd2020nonlinear}. In the case of a medium stationary in Schwarzschild' spacetime, by employing the correspondence with an \emph{effective medium} in flat spacetime as described in the previous section, the usual derivation of the NLSE can be carried out. However, the effective medium will be inhomogeneous due to the curved spacetime contribution to the polarizability and permeability of the material medium. This gives rise to extra terms in the NLSE which are of purely gravitational origin. Furthermore, another source of inhomogeneity in the medium can be included when considering the effect of tidal forces on the material that, through photoelasticity, render the refractive index position-dependent. 

Neglecting for the moment photoelasticity, i.e., considering a rigid dielectric, we can write Maxwell's equation in flat spacetime for the effective medium in the familiar notation, {using the fields and field strengths that we indicate with plain capital letters from now on,}
\begin{align}\label{Meq}
    & \nabla\cdot B=0,\,\,\nabla\cdot D=0\\
    & \nabla\times E=-\partial_t B,\,\, \nabla\times H=\partial_t D,
\end{align}
where $D=\tilde{\varepsilon} E$ and $H=B/\tilde{\mu}$. Here $\tilde{\mu}=\tilde{\mu}(r)$ and $\tilde{\varepsilon}=\tilde{\varepsilon}(E,r,\omega)$ in frequency space, allowing us to account for the effect of material dispersion, are the permeability and permittivity of the effective medium. Expressing the 
Schwarzschild' spacetime metric in isotropic coordinates as $ds^2=-\left(B(t,r)/A(t,r)\right)^2 dt^2+A^4(t,r)\delta_{\alpha\beta}dx^{\alpha}dx^{\beta}$, with $A(r)=1+r_S/4r$ and $B(r)=1-r_S/4r,$ with $r_S$ the Schwarzschild radius, we have 
\begin{align}\label{eq:emu}
    &\tilde{\varepsilon}(E,r,\omega)=\varepsilon_0\varepsilon_{\rm sp}\varepsilon=\varepsilon_0 \frac{A(r)^3}{B(r)}\left(1+\chi^{(1)}(\omega)+3\chi^{(3)}\frac{|E|^2}{\Omega}\right),\\
    &\tilde{\mu}=\tilde{\mu}(r)= \mu_0{\mu_{\rm sp}=\mu_0} A(r)^3B(r)^{-1},
\end{align} 
with $\Omega=A(r)^{-4}$ the conformal factor relating the spacial part of the metric with the flat, Euclidean one\footnote{This conformal factor arises due to the fact that $E^aE_a$ in {curved} spacetime corresponds to $|E|^2/\Omega$ with $|E|^2=E^aE^b\delta_{ab}$ the flat spacetime norm squared of the electric strength field.}.
The explicit radial dependence in the linear part of these effective quantities comes from the curved spacetime optical properties encoded in {the diagonal terms $\sqrt{-g}g^{\alpha\alpha}/g_{00}$ (cf. eq.\eqref{permten}) that we define as} $\varepsilon_{\rm sp}=\mu_{\rm sp}=A(r)^3B(r)^{-1}$. The field dependency of $\tilde{\varepsilon}$ takes into account the non-linearity of the physical medium. Note also that dispersion implies that the dielectric permeability is a function of the physical frequency $\omega$ defined with respect to our stationary observer $u^\mu$. 

From eq.~\eqref{Meq}, and writing $D=\tilde{\varepsilon}_\ell E+P_{\rm NL}$, where {$\tilde{\varepsilon}_\ell=\varepsilon_0\varepsilon_{\rm sp}(1+\chi^{(1)}(\omega))$} is the linear part of the dielectric permeability {in eq.~\eqref{eq:emu}} and $P_{\rm NL}$ is the non-linear polarization, we can then obtain the wave equation, in frequency space,
\begin{equation}\label{start}
 \nabla^2 E-\nabla(\nabla\cdot E)+\tilde{\mu}\tilde{\varepsilon}_\ell \nu^2E=-\tilde{\mu} \nu^2 P_{\rm NL}-(\nabla\log({\mu_{\rm sp}}))\times(\nabla\times E)\,.
\end{equation}
Here we indicate with $\nu$ the conjugate variable to the coordinate time $t$ in the flat spacetime of the effective medium. 
Note that the homogeneous Maxwell equations imply that 
\begin{align}\label{homoeq}
    \nabla \cdot E  =   -  (\nabla \log\tilde\varepsilon_\ell)\cdot E - \frac{1}{\tilde\varepsilon_\ell}\nabla \cdot P_{\rm NL},
\end{align}
and thus 
\begin{align}\label{homo2}
    - \nabla(\nabla\cdot E) &= (E\cdot\nabla)\nabla \log\tilde\varepsilon_\ell + \left((\nabla\log\tilde\varepsilon_\ell)\cdot \nabla\right)E \\ \nonumber
    &+ (\nabla\log\tilde\varepsilon_\ell)\times(\nabla\times E)+\nabla\left(\frac{1}{\tilde\varepsilon_\ell} (\nabla\cdot P_{\rm NL})\right).
\end{align}
Eq.~\eqref{homoeq} makes evident that $\nabla \cdot E$ is of the same order as the non-linearities and inhomogeneities in the electric permittivity, which is also why it is usually safely neglected in derivations of the NLSE.

The wave equation in eq.~\eqref{start} is equivalent to Maxwell equations and, as such, presents the same level of complexity if analytical or numerical solutions are attempted. 
The NLSE is a scalar propagation equation for the electric field's slowly varying amplitude that allows 
one 
to numerically simulate the pulse propagation. We thus want to write the electric field as the product of a slowly varying amplitude times a phase propagating along the propagation direction, that we will identify with the $z$ direction in the following. 
In this context, notice that the dispersion relation of the physical medium, in its rest frame, {is given simply by} $n(\omega)=c{\kappa}/{\omega}$, with $\kappa$ the modulus of the spatial projection of the wave 4-vector. For the effective medium, this relation reads $\tilde{n}=c{\tilde{\kappa}}/{\nu}$, where $\tilde{n}=\sqrt{\varepsilon_{\rm sp}\mu_{\rm sp}}n$ is the product of the material refractive index and the ``spacetime refractive index'' $n_{\rm sp}=\sqrt{\varepsilon_{\rm sp}\mu_{\rm sp}}$. Moreover, since $\nu$ is the frequency defined with respect to Minkowski coordinate time, i.e., the conjugate Fourier variable to $t$, it is related to the physical frequency, {i.e., the one measured by a physical observer in curved spacetime,} by the gravitational redshift $\nu=\omega\sqrt{-g_{00}}$. From the equivalence of the dispersion relations, we see that $\tilde{\kappa}(r)=\kappa n_{\rm sp}(r)\sqrt{-g_{00}(r)}$.  
We will thus write $E({\bf r},t)\propto \mathcal{E}({\bf r})e^{i(\tilde{\kappa}_0 z-\nu_0 t)}+cc.$, with $\kt_0=\kt(r,\nu_0)$ evaluated at a central frequency $\nu_0$.

In order to proceed with the derivation of the NLSE, and to further simplify our equations, we consider two separate situations of physical interest: (i) pulse propagation at approximately constant radius; (ii) pulse propagating radially. 

\subsubsection{Horizontal motion at (almost) constant radius}
We assume the propagation direction of the light pulse to be the $z$ axis 
taken to be perpendicular to the radial direction for horizontal motion, and consider linearly polarized light propagating in a medium stationary on Earth for concreteness. Then, for propagation distances much smaller than Earth's {radius ($r_\oplus$), i.e., $z\ll r_\oplus$, the horizontal motion can be considered as happening at constant radius.}
With these approximations, the spacetime permeability and permittivity are constant functions of $r_\oplus$, $\mu_{sp}=\varepsilon_{\rm sp}=A(r_\oplus)^3B(r_\oplus)^{-1}$ and also the physical frequency is not changing with $z$. Thus, we see that in eq.~\eqref{start} the last term on the right-hand side vanishes. 

We follow the derivation in~\cite{philbin2008fiber} where the pulse propagation in a single-mode optical fiber was considered. Indeed, {for $\mu_{sp}\,\varepsilon_{\rm sp}$ constant,} eq.~\eqref{start} is formally equivalent to eq.~(S1) of~\cite{philbin2008fiber} {in frequency space}. We thus end up with an effective one dimensional problem for the slowly varying envelope, and the derivation of the NLSE is the textbook one~\cite{SM,boyd2020nonlinear}. In particular, recall that the slowly varying envelope approximation(s) (SVEA) consists in neglecting terms $\partial^2_z \mathcal{E}\ll\kt_0\partial_z \mathcal{E}$ and $(\kt_1/\kt_0)\partial_t\ll 1$ on the basis that the envelope will contain many wavelengths and optical cycles. If we apply now the SVEA we end up with, in the time domain, 
\begin{align}
&i(\partial_z +\kt_1\partial_t)\mathcal{E}
-\frac{\kt_2}{2} \partial^2_t \mathcal{E} ={-n_2\nu_0 n_{\rm sp}(r_\oplus) \varepsilon_0 \frac{|\mathcal{E}|^2}{\Omega} \mathcal{E}},\label{eq:soliton_diffeq_hor1}
\end{align}
where $\kt_i(\nu_0)$ are the coefficients of the power series expansion $\kt(\nu)= \sum_{n}\tilde{\kappa}_n(\nu_0)/n!\,(\nu-\nu_0)^n$ in $\nu-\nu_0$ and we are considering Kerr non-linear media for which the nonlinear index is $n_2=3\chi^{(3)}/(2n(\omega_0)c\varepsilon_0)$.

{Considering an anomalous dispersive material, i.e., $\kappa_2(\nu_0)<0$, an analytical solution of the NLSE can be found {(see, e.g.,~\cite{philbin2008fiber})} 
 and reads}
\begin{equation}\label{eq:soliton_horizontal_main}
\mathcal{E}(t,z)=\sqrt{\frac{\Omega | \kt_2| }{\nu_0 n_2 n_{\rm sp}\varepsilon_0 T_0^2 }}\cosh \left(\frac{t-\kt_1 z}{T_0}\right)^{-1}\exp \left(\frac{i z | \kt_2| }{2 T_0^2}\right),
\end{equation}
where $T_0$ is the pulse length, and $1/\kt_1$ is its speed of propagation. This reduces to the result from Philbin et al.\cite{philbin2008fiber} {-- eq.(S74) of the supplementary material in~\cite{philbin2008fiber} --} in the limit of $r_S\to 0$. From this expression, combined with the fact that $\kt_1(\nu_0)=n_{\rm sp}\kappa_1(\omega_0)$, we can conclude that the {velocity of the horizontally propagating soliton in curved spacetime with respect to an observer comoving with the segment of the dielectric material\footnote{{Indeed note that proper length and proper time for an observer comoving with the segment of the dielectric material and in connection with coordinate quantities 
are given by $\ell=A^2\,z$ and $\tau=t\,B/A$ so that $v\equiv \ell/\tau=A^3B^{-1}z/t=n_{\rm sp}\tilde{v}$.   }}} 
is given simply by $\kappa_1(\omega_0)^{-1}$. 
\begin{figure}
\centering
\includegraphics[scale=0.4]{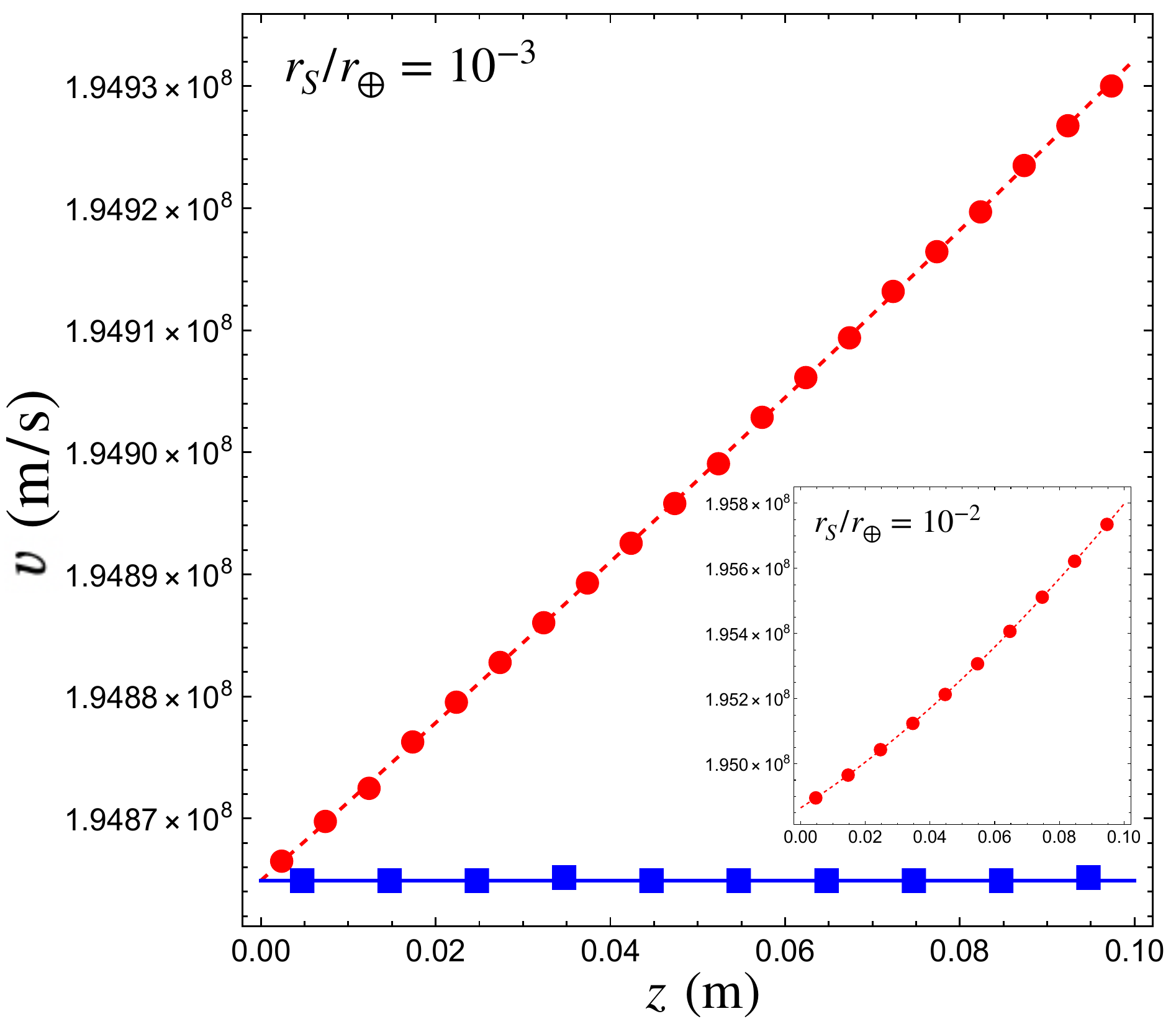}
    \caption{Velocity of the soliton along the fiber, {with respect to an observer comoving with the segment of the dielectric material where the (peak of the) soliton is located,} for $L=0.1$\,m, $r_s=10^{-3} r_\oplus$, and including photoelasticity. The red, dashed and blue, solid curves represent the analytical expression in eq.~\eqref{eq:analytical_v} including or in the absence, respectively, of photoelasticity. The red points and blue squares are obtained by numerical simulations and 
    agrees 
    perfectly with the analytical formula of eq. \eqref{eq:analytical_v}. The inset shows the case with photoelasticity in which $r_s=10^{-2} r_\oplus$. This shows a deviation from a purely linear relation between the velocity and the propagation distance. 
    }
    \label{fig:velocityofz}
\end{figure}
\subsubsection{Radial motion}
Let us now consider the case in which the light pulse propagates radially along the $z$ direction. Care is in order here, since now all the quantities appearing in the wave equation will change along the propagation direction, including the physical frequency that will be subject to gravitational redshift. Motivated by the symmetry of the problem, and in order to obtain a scalar, one-dimensional equation whose solution can be simulated, we assume that all the quantities entering the wave equation depend solely on $z$. This is tantamount to identifying the radial direction with the $z$-axis {and work close to $x=y= 0$} so that $r= r_\oplus+z$, which is a reasonable assumption since we are considering the vertical propagation of a well localized pulse. With this approximation, the wave equations~\eqref{start} reduce to a system of three decoupled equations~\cite{habib2013electromagnetic} 
{\small
\begin{align}
   \partial_z^2E_{x(y)}+\tilde{\mu}\tilde{\varepsilon}_\ell\nu^2 \label{eq:zdirection1} E_{x(y)}=&-\tilde{\mu}\nu^2P_{{\rm NL,} x(y)}+\left(\partial_z(\ln\tilde{\mu})\right)\partial_z E_{x(y)}
\end{align}
\begin{align}
   \partial_z^2E_{z}+\tilde{\mu}\tilde{\varepsilon}_\ell\nu^2 E_{z}=&-\tilde{\mu}\nu^2P_{{\rm NL,} z}-\partial_z\left(\frac{1}{\tilde{\varepsilon}_\ell}\partial_z P_{{\rm NL},z}\right)\\ \nonumber
   &-2(\partial_z\ln\tilde{\varepsilon}_\ell)\partial_zE_{z}-E_z\partial_z^2\ln\tilde{\varepsilon}_\ell 
\end{align}
}
It is immediate to realize that $E_z=0$ is a solution of the corresponding equation so that we can consider the propagation of linearly polarized light (in a direction orthogonal to $z$) and we end up with a single equation of the form of eq.~\eqref{eq:zdirection1}.

Proceeding as before with substituting the ansatz $E(z,t)\propto \mathcal{E}(z,t)e^{i(\tilde{\kappa}_0(z) z-\nu_0 t)}+cc.$, expanding $\kt(z,\nu)$ around $\nu_0$, and using the SVEA approximation(s) we obtain the NLSE given by 

\begin{widetext}
{\small
\begin{align}
&i(\partial_z +\kt_1\partial_t)\mathcal{E}-\frac{\kt_2}{2} \partial^2_t \mathcal{E}  +2i\frac{\partial_z\kt_0}{2\kt_0} \mathcal{E}+ 2i z\frac{\partial_z\kt_0}{2\kt_0}\partial_z \mathcal{E}+i z\frac{\partial_z^2\kt_0}{{2\kt_0}} \mathcal{E}-z\partial_z\kt_0 \mathcal{E}-z^2\frac{(\partial_z\kt_0)^2}{2\kt_0} \mathcal{E}={-n_2\nu_0 n_{\rm sp}(r) \varepsilon_0 |\mathcal{E}|^2 \mathcal{E}/\Omega} +\frac{\partial_z\ln n_{\rm sp}}{2\kt_0}\left(i\kt_0\mathcal{E}+\partial_z \mathcal{E}+iz(\partial_z\kt_0) \mathcal{E}\right).\label{eq:soliton_diffeq_vertical}
\end{align}
}
\end{widetext}
Eq.~\eqref{eq:soliton_diffeq_vertical} contains several additional terms with respect to the equation for the horizontal propagation due to the fact that now the wavevector $\kt_0$ depends explicitly on the 
coordinate along the 
propagation direction and so does the refractive index, i.e., we are propagating in a gradient-index medium (GRIN)\footnote{{See also~\cite{PhysRevLett.37.693,chen1978nonlinear,herrera1984envelope} for early studies of soliton propagation in inhomogeneous  media.}}. 
All geometrical quantities appearing in the equation are evaluated at $r_\oplus+z$. Finally, consistently with the horizontal propagation case, upon setting $\kt_0$ constant, we return to eq.~\eqref{eq:soliton_diffeq_hor1}.

\section{Including photoelasticity}
Up until now, we have considered rigid dielectrics, i.e., dielectric media in which the speed of sound is infinite. For realistic materials, this is of course never the case and the dielectric gets deformed by the action of forces, including the tidal ones in our set-up. Let us consider an optical fiber as a paradigmatic example. In this case, the deformation due to the action of gravity will be relevant only for the case of vertical propagation.

Deformations of a dielectric lead to a change in the relative permeability of the material, and thus of the refractive index, a phenomenon known as photoelasticity~\cite{chen2006foundations}. The contributions to this effect coming from the curvature of spacetime and the inertial acceleration of the fiber can be separately accounted for following the discussion in~\cite{ratzel2018frequency}. Consider a fiber of length $L$ hanging from at support located at $r_\oplus+L$. As far as the strain is within the elastic limit of the material, we can relate it with the stresses through a linear relation, i.e., Hooke's law. Thus, we write the strain tensor as $\mathcal{S}_{kl} =\frac{1}{Y} \sigma_{kl}$, where $Y$ is the Young's modulus of the material and $\sigma_{kl} = \frac{F_k}{A_l}$ is the stress tensor given by the ratio between the force $F_k$ in direction $\hat{e}_k$ and the cross-sectional area $A_l$ normal to $\hat{e}_l$ upon which the force acts. The photoelastic (or acousto-optic) effect consists in the change of the relative electric permeability by $\Delta(\bm{\varepsilon}_r)^{-1}_{kl} = \mathcal{P}_{kl\,mn}\mathcal{S}_{mn}$, where $\mathcal{P}$ is the photoelastic tensor. In the following, we limit ourselves to the case of isotropic materials and a diagonal stress tensor (see~\cite{SM} for the details of the computation).
It should be noted that photoelasticity is far from negligible in the case under investigation and becomes the dominant effect in the vertical propagation scenario,  overwhelming 
the effect related to the optical properties of the background spacetime. 

While photoelasticity introduces a further radial dependence in the optical properties of the effective medium, this does not affect the form of eq.~\eqref{eq:soliton_diffeq_vertical}, which remains valid. The only difference is in the expressions for the quantities $\kt_i$ and their derivatives, due to the fact that now the refractive index of the medium is given by $n(\omega)=\sqrt{1+\chi_1(\omega)+\Delta{\varepsilon}_r(\omega)}$~\cite{SM}. 
\begin{figure}
\centering
\includegraphics[scale=0.45]{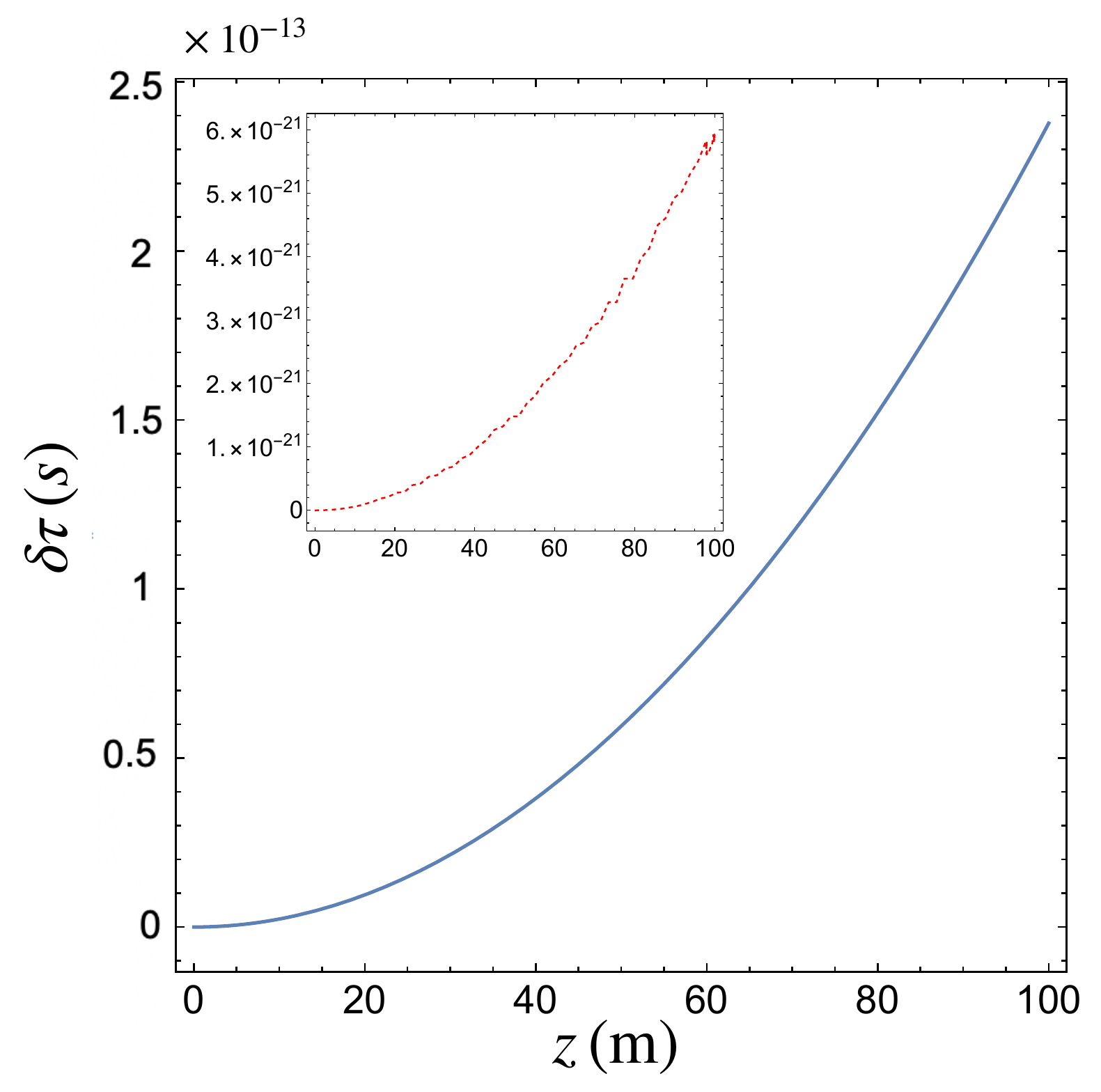}
    \caption{Time of arrival of the soliton for the case of propagation in the gravitational field of Earth {for which we assume} $r_S = 9\times 10^{-3}$\,m. The main figure shows the difference in time of arrival, {with respect to an observer comoving with the segment of the dielectric material where the soliton is located,} between vertically and horizontally propagating solitons over the propagation coordinate length $z$. The inset shows the same in the case photoelasticity is neglected. 
    }
    \label{fig:arrivaltime}
\end{figure}

\section{Numerical results}
While the wave equation in eq.~\eqref{start} gives us the full Maxwell equations, including possibly interesting effects related to the vectorial nature of the electric field, and thus to the interplay between gravity and the light polarization, its numerical investigation is beyond the scope of the current work, and it is left for future investigations. Here, we focus on the propagation of light pulses as described by the simplified eq.\eqref{eq:soliton_diffeq_vertical}, motivated by light propagation in optical fibers~\cite{philbin2008fiber}. Note that in the case of eq.~\eqref{eq:soliton_diffeq_hor1} an analytical solution was presented in eq.~\eqref{eq:soliton_horizontal_main}.

Equation \eqref{eq:soliton_diffeq_vertical} for the vertical propagation is solved numerically -- being a non-linear PDE with coordinate dependent coefficients -- using the split-step Fourier (SSF) method~\cite{agrawal2000nonlinear} and taking into account also the effect of the fiber deformation. 
For this purpose, we utilize the same fiber parameters as in~\cite{philbin2008fiber} (see also table~\ref{tab:parameters} in~\cite{SM}) and initialize the temporal profile at $z=0$ as the one of the input pulse in the same reference. 

The intuition based on the SSF method{-- where the propagation equation~\eqref{eq:soliton_diffeq_vertical} is rewritten in the form $\partial_z \mathcal{E}= \left(\hat{D}+\hat{N}\right)\mathcal{E}$ with the diffusive dynamics enclosed in the operator $\hat{D}=\hat{D}(z,\partial_t)$~\cite{SM} --} allows us to {formulate the educated guess}
that the propagation speed of the soliton, in the effective flat spacetime, is given by 
\begin{equation}\label{eq:analytical_v}
    \tilde{v}=\frac{1+z\, \kt_0'(z)/\kt_0(z)}{\kt_1(z)}.
\end{equation}
Indeed, this appears as {(the real part of) the inverse of} the coefficient of the time derivative {in $\hat{D}(z,\partial_t)$}. Then, in order to translate this result {into the speed measured by an observer comoving with the segment of the dielectric material where the soliton peak is located,} we need to just multiply eq.~\eqref{eq:analytical_v} by the spacetime refractive index. That this intuition is indeed correct is verified by the numerical simulations reported in Fig.~\ref{fig:velocityofz}. 
We see that the $z$-dependence of the propagation velocity is strongly enhanced by the effects of mechanical deformation of the fiber {with respect to the case in which photoelasticity is ignored.} 
{The $z$-dependence of the vertical propagation velocity without photoelasticity is weak, and the velocity is close to the one of the horizontal case}. To quantify the latter statement, in Fig.~\ref{fig:arrivaltime} we show the difference in the (proper) time of arrival of the soliton for the case of propagation in the gravitational field of Earth, corresponding to 
a Schwarzschild radius that we take as
$r_S= 9\times 10^{-3}$~m. The main figure shows 
{\begin{equation}
    \delta \tau=|z(\sqrt{-g_{00}(r_\oplus+z)} \tilde{v}_{\uparrow}^{-1})- \sqrt{-g_{00}(r_\oplus)} \tilde{v}_{\rightarrow}^{-1})|,
    \end{equation}
    with $\tilde{v}_{\uparrow}$ and $\tilde{v}_{\rightarrow}$}
the propagation velocities, in the effective flat spacetime, for vertical and horizontal propagation. The inset shows instead the case in which for the vertical propagation the photoelasticity is neglected, showing a much weaker dependence.

\begin{figure}
\includegraphics[scale=0.45]{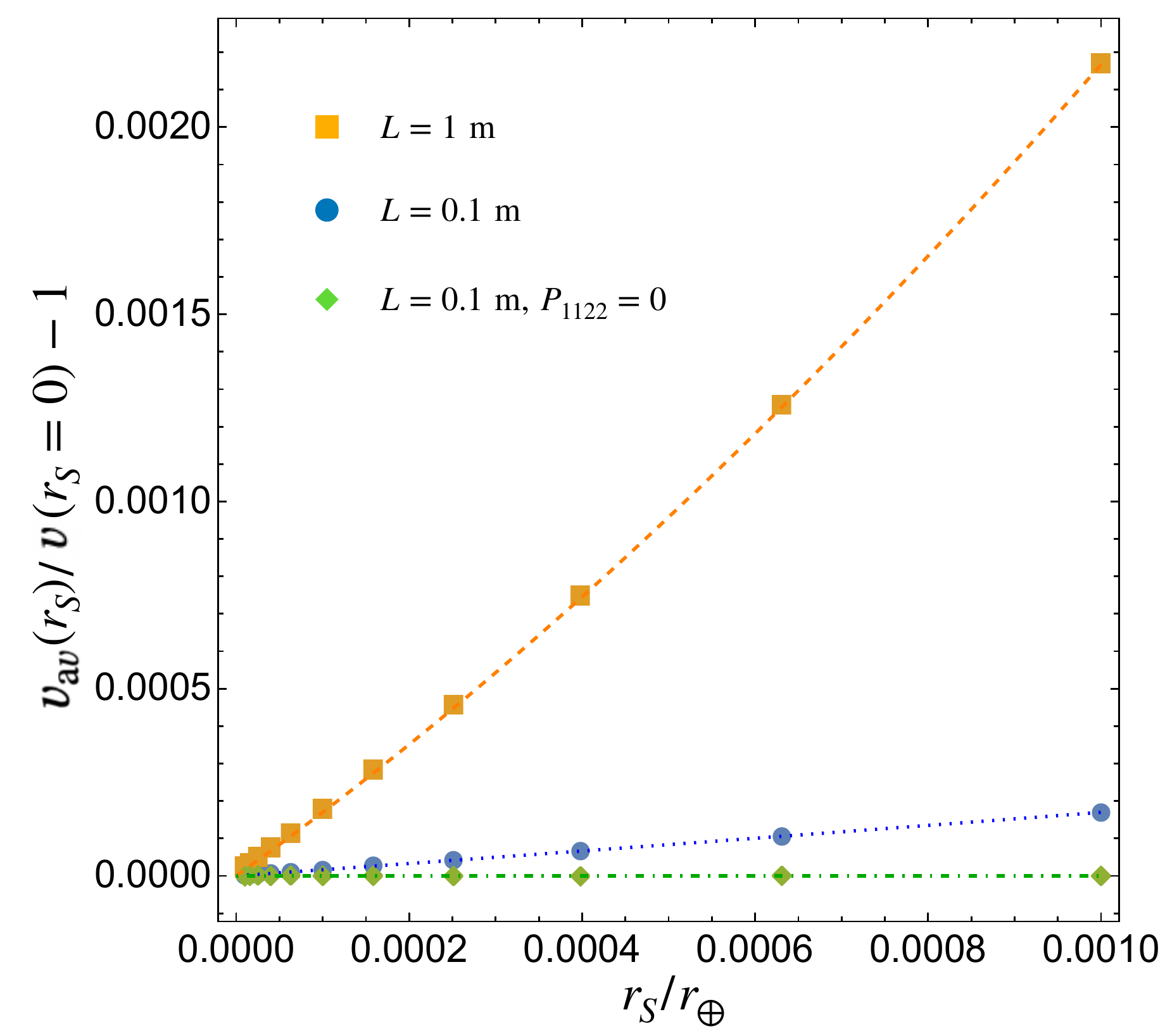}
    \caption{Change in average velocity ($v_{\rm av}$) of the soliton in the fiber -- {with respect to the observer comoving with the dielectric --} 
    {compared to the case with $r_S=0$}. Orange, square points corresponds to the case of a $L=1$\,m propagation with photoelasticity. Blue, round points correspond to the case of a $L=0.1$\,m propagation with photoelasticity. Green,  diamonds correspond to the case of a $L=0.1$\,m propagation without photoelasticity. The lines correspond to the analytical result that fits perfectly the different sets of data. 
    }
    \label{fig:avgvelocity}
\end{figure}

{Finally, in Fig.~\ref{fig:avgvelocity} we show the deviation of the average velocity along the vertical direction $v_{\rm av}(r_S)$ with respect to the constant velocity at $r_S=0$ as a function of the dimensionless ratio $r_S/r_\oplus$. The average velocity is obtained numerically from the simulations as the ratio of the total length $L$ and the propagation time of the soliton and transformed into the frame of the {observer comoving with the fiber at its upper end-point} -- i.e., multiplied by $n_{\rm sp}(r_\oplus+L)$. Analytically, we use $v_{\rm av}=(\int_{0}^{L}v\,{\rm d}z)/L$ with $v=n_{\rm sp}\tilde{v}$ and $\tilde{v}$ given in eq.~\eqref{eq:analytical_v}. Fig.~\ref{fig:avgvelocity} shows once again the agreement between the simulated data and our analytical ansatz and it also shows that the photoelasticity is the main effect that allows one to have a sizable difference between the flat and curved spacetime propagation.}

{Another quantity characterizing the propagating pulse is its temporal width. In the horizontal propagation case, the duration of the pulse is constant. The same is not, in general, true when considering the vertical propagation. In the Supplemental Material~\cite{SM}, we report the evolution of the temporal width along the fiber. In particular, our simulations show a focusing of the pulse which is however sizable only in the presence of photoelasticity.}

\section{Conclusions}
We have considered the propagation of light pulses in non-linear, non-magnetic media stationary in curved spacetime. Taking some intuition from the seminal work of Plebanski~\cite{plebanski1960electromagnetic}, we 
showed that light propagation in such media can be equivalently described as the propagation in an effective medium in flat spacetime whose electric and magnetic properties acquire a multiplicative factor encoding the spacetime structure. Having done that, eq.~\eqref{start} describes the propagation of light in the effective medium. 
It is interesting to note, even though we did not investigate it in this work, that the vectorial nature of this equation encodes the interplay between the light polarization and the gravitational field. Such interplay should be expected on the basis of the fact that the effective medium is an inhomogeneous, gradient-index medium for which it is well known that the propagation of light is influenced by its own polarization~\cite{bliokh2009geometrodynamics,liberman1992spin,bliokh2015spin}. Furthermore, the effect of polarization  
on the propagation of light in curved, vacuum spacetime has been extensively considered in the literature and shown to take place also 
for static spacetimes~\cite{gosselin2007spin,oancea2020gravitational}.

Neglecting the aforementioned effects, which would be undoubtedly small, by virtue of approximations we have been able to derive a scalar NLSE describing the propagation of a light pulse. It is important to notice that, when solving the NLSE employing the SSF method, we are implicitly considering a unidirectional equation and {ignoring any possible back-propagating field in the boundary conditions imposed, for all times, at $z=0$. This means that backscattered light from the pulse is assumed negligible relative to the
pulse itself, a condition common to all unidirectional envelope propagation equations~\cite{PhysRevLett.89.283902}}.
While this is not a problem for the horizontal propagation, in which case only the weak non-linearity could give rise to back-reflection, in the case of the vertical propagation light is effectively propagating in a gradient-index medium with the refracting index slowly varying in the propagation direction. This by itself can give rise to back-{propagating fields}, and effectively limits the validity of our treatment to regimes in which the photoelasticity {allows to employ a unidirectional equation.}
Luckily, the regime of validity of the equation -- which depends on the parameter chosen for the physical medium -- can be readily estimated by following the discussion in~\cite{PhysRevA.81.013819} as we detail in~\cite{SM}.

{Given these caveats, the NLSE that we have derived shows that an optical pulse propagating radially in a Kerr non-liner medium stationary in Schwarzschild spacetime experiences a change in its propagation velocity captured by eq.~\eqref{eq:analytical_v}. This effect is mostly due to photoelasticity which overwhelms the purely spatiotemporal effects encoded in $n_{\rm sp}$. The difference in propagation velocity between the vertically and horizontally propagating pulses results, in turn, in a difference of the time of arrival of two pulses of the order of hundreds of femtoseconds in Earth gravitational field,}{ a fact that puts this difference in the reach of current technologies (see \cite{lee2010time,fortier201920,caldwell2022time} and references therein).}

\section*{Acknowledgements}
The authors thank Francesco Marino for interesting discussions. A.~Belenchia and D.~Braun acknowledge  support  from the Deutsche Forschungsgemeinschaft (DFG, German Research Foundation) project number BR 5221/4-1.  {D.~R\"{a}tzel acknowledges funding by the Federal Ministry of Education and Research of Germany in the project “Open6GHub” (grant number: 16KISK016) and support through the Deutsche Forschungsgemeinschaft (DFG, German Research Foundation) under Germany’s Excellence Strategy – EXC-2123 QuantumFrontiers – 390837967, the Research Training Group 1620 “Models of Gravity” and the TerraQ initiative from the Deutsche Forschungsgemeinschaft (DFG, German Research Foundation) – Project-ID 434617780 – SFB 1464.}

\bibliography{references2.bib}

\widetext
\clearpage
\begin{center}
\textbf{\large Supplemental Material: Optical solitons in curved spacetime}
\end{center}
\setcounter{equation}{0}

\setcounter{figure}{0}
\setcounter{table}{0}
\setcounter{page}{1}
\makeatletter
\renewcommand{\theequation}{S\arabic{equation}}
\renewcommand{\thefigure}{S\arabic{figure}}

\begin{center}
  Felix Spengler$^{1}$, Alessio Belenchia$^{1,2}$, Dennis R\"{a}tzel$^{3}$, Daniel Braun$^1$ \\ \vspace{0.2cm} \small{
  {\em $^1$Institut f\"{u}r Theoretische Physik, Eberhard-Karls-Universit\"{a}t T\"{u}bingen, 72076 T\"{u}bingen, Germany}\\
{\em $^2$Centre for Theoretical Atomic, Molecular, and Optical Physics,\\School of Mathematics and Physics, Queen’s University, Belfast BT7 1NN, United Kingdom}\\
{\em $^3$Humboldt  Universit\"{a}t  zu  Berlin,  Institut  f\"{u}r  Physik, Newtonstraße  15,  12489  Berlin,  Germany}\\
}
\end{center}

In this supplemental material, we collect the detailed derivations of the results in the main text.

\section{Vacuum spacetime as an optical medium \& the \emph{effective medium} description}
Thanks to the seminal work of Plebanski in the '60s~\cite{plebanski1960electromagnetic}, it is well known that electromagnetism in curved spacetime is equivalent to propagation in an optical medium. Following the derivation presented in~\cite{de1971gravitational}, Maxwell vacuum equations in curved spacetime are written as 
\begin{align}
    \nabla_k F^{*\,ik}=0\label{homo}\\
    \nabla_k F^{ik}=0,\label{nohomo}
\end{align}
where $F^{*}$ is the Hodge dual of the e.m. tensor, {Latin indices run from 0 to 3, and the metric $g_{ij}$ has mostly plus signature}. As in the main text, we consider the case with no currents. 

Choosing an observer field $u^i$, the electric and magnetic field strength can be defined with respect to it as 
\begin{align}
   & H^i=F^{*\,ij}u_j,\,\,E_i=F_{ij}u^j\\
    &F_{ij}=\eta_{ijkl}u^l H^k+2u_{[i}E_{j]},
\end{align}
where here $\eta_{ijkl}=\sqrt{-g}\delta_{ijkl}$ is the Levi-Civita tensor (with $\delta_{ijkl}$ the Levi-Civita alternating symbol in four dimensions) and $T_{[abc\dots]}$ denotes the antisymmetrization of the tensor with respect to the indices in square brackets.
The Maxwell equations can then be projected in the $u^{i}$ direction or orthogonal to it using the projection operator into the rest frame of $u^{i}$, $h_{ij}=g_{ij}+u_i u_j$. 
The end result is, in the case the observer field is chosen as $u^{i}=\delta^i_0/\sqrt{-g_{00}}$
\begin{align}
    &\delta^{\alpha\beta\gamma}\partial_\beta\mathcal{H}_\gamma-\partial_0\mathcal{D}^\alpha=0;\,\,\partial_l\mathcal{D}^l=0\\
    &\delta^{\alpha\beta\gamma}\partial_\beta\mathcal{E}_\gamma+\partial_0\mathcal{B}^\alpha=0;\,\,\partial_l\mathcal{B}^l=0,
\end{align}
where the first two equations come from Maxwell equations~\eqref{nohomo} (with $\delta^{\alpha\beta\gamma}$ the Levi-Civita alternating symbol in three dimensions) while the second two from eq.~\eqref{homo}. Here, $\mathcal{H}_\alpha=\sqrt{-g_{00}}H_\alpha$, $\mathcal{E}_\alpha=\sqrt{-g_{00}}E_\alpha$, Greeks indices run from 1 to 3, and 
\begin{align}
    & \mathcal{D}^\alpha=-\sqrt{-g}\frac{g^{\alpha\beta}}{g_{00}}\mathcal{E}_\beta-\delta^{\alpha\beta\gamma}\frac{g_{0\gamma}}{g_{00}}\mathcal{H}_\beta\\
     & \mathcal{B}^\alpha=-\sqrt{-g}\frac{g^{\alpha\beta}}{g_{00}}\mathcal{H}_\beta+\delta^{\alpha\beta\gamma}\frac{g_{0\gamma}}{g_{00}}\mathcal{E}_\beta.
\end{align}
From here one can see that these equations are actually equivalent to Maxwell equations in flat spacetime in the presence of an optical medium whose constitutive relations are characterized by a dielectric {($\varepsilon_{\rm sp}^{\alpha\beta}$) and magnetic permeability ($\mu_{\rm sp}^{\alpha\beta}$)} given by 
\begin{align}
    & \mu_{\rm sp}^{\alpha\beta}=\varepsilon_{\rm sp}^{\alpha\beta}=-\sqrt{-g}\frac{g^{\alpha\beta}}{g_{00}}.
\end{align}

As shown in the main text, when a physical optical medium whose rest frame is characterized by $u^i$ is added, we can follow the same derivation starting from Maxwell's equations in curved spacetime and with a material medium 
\begin{align}
    \nabla_k F^{*\,ij}=0\\
    \nabla^k G_{ik}=0, 
\end{align}
where now $G$ and $F$ are related by the material constitutive equations. 
Then we have 
\begin{align}
    & B_a=-\frac{1}{2}\eta_{abcd}u^b F^{cd};\,\,E_i=F_{ij}u^j\\
    & H_a=-\frac{1}{2}\eta_{abcd}u^b G^{cd};\,\,D_i=G_{ij}u^j\\
    &F_{ab}=-\eta^{cd}_{ab}u_d B_c+2u_{[a}E_{b]}\\
    &G_{ab}=-\eta^{cd}_{ab}u_d H_c+2u_{[a}D_{b]},
\end{align}
where we have introduced the electric and magnetic excitation, $D_a$ and $H_a$ respectively, on top of the electric and magnetic strength $E_a$ and $B_a$.

Note that the definitions of $E,B, F$ are equivalent to the vacuum case, since the homogeneous Maxwell equations are the same. The inhomogeneous equations have also the same form as in the vacuum case, but with the substitution of $E,B,F$ with $D,H,G$, where the definition of $G$ with respect to $H, D$ is the same as $F$ with respect to $E,B$. 
From this simple observation we can immediately deduce that the projection of Maxwell equations in 3-dimensional form will, in the case the observer field is chosen as $u^{i}=\delta^i_0/\sqrt{-g_{00}}$, lead to
\begin{align}
    &\delta^{\alpha\beta\gamma}\partial_\beta\mathcal{H}_\gamma-\partial_0\mathcal{D}^\alpha=0;\,\,\partial_l\mathcal{D}^l=0\\
    &\delta^{\alpha\beta\gamma}\partial_\beta\mathcal{E}_\gamma+\partial_0\mathcal{B}^\alpha=0;\,\,\partial_l\mathcal{B}^l=0,
\end{align}
where $\mathcal{E}_\alpha=\sqrt{-g_{00}}E_\alpha$, $\mathcal{H}_\alpha=\sqrt{-g_{00}}H_\alpha$, and
\begin{align}\label{Maxelleff2}
    & \mathcal{D}^\alpha=-\sqrt{-g}\frac{g^{\alpha\beta}}{g_{00}}\mathfrak{D}_\beta-\delta^{\alpha\beta\gamma}\frac{g_{0\gamma}}{g_{00}}\mathcal{H}_\beta\\
     & \mathcal{B}^\alpha=-\sqrt{-g}\frac{g^{\alpha\beta}}{g_{00}}\mathfrak{B}_\beta+\delta^{\alpha\beta\gamma}\frac{g_{0\gamma}}{g_{00}}\mathcal{E}_\beta,
\end{align}
with $\mathfrak{B}_\alpha=\sqrt{-g_{00}}B_\alpha$, and $\mathfrak{D}_\alpha=\sqrt{-g_{00}}D_\alpha$.
{
Once again, these equations are equivalent to Maxwell's equations in flat spacetime in the presence of an effective optical medium. 

Consider the case of a linear, dispersionless medium. We can then write $G^{ij} = \frac{1}{2}\chi^{ij\,kl} F_{kl}$, with the material's constitutive tensor $\chi^{ij\,kl}$, containing all material properties, which is symmetric under the exchange of the first and second pair of indices and antisymmetric with respect to the swap within an index pair. In particular, we can also write $D_a=\varepsilon^b_a E_b$, and $B_a=\mu^b_a H_b$, which are the constitutive relations in the reference frame of the observer in which the medium is at rest, neglecting magneto-electric effects. For an isotropic medium, we also have that the dielectric and permeability tensor assume the simplified form $\varepsilon^b_a=\varepsilon(\delta^b_a+U^bU_a)$ and $\mu^b_a=\mu(\delta^b_a+U^bU_a)$ for some scalar, positive functions $\varepsilon$ and $\mu$. The effective optical medium is such that its constitutive relations are then characterized by a dielectric and inverse magnetic permeability given by
\begin{align}
    & \tilde{\varepsilon}^{\alpha\beta}=-\sqrt{-g}\frac{g^{\alpha\gamma}}{g_{00}}\varepsilon_\gamma^{\;\beta},\\
    & \tilde{\mu}^{\alpha\beta}=-\sqrt{-g}\frac{g^{\alpha\gamma}}{g_{00}}\mu_\gamma^{\;\beta},
\end{align}
while the antisymmetric parts of the constitutive tensor are completely characterized by the vacuum spacetime properties\footnote{{More in general, one could also include in this description materials for which the magnetoelectric entries of the constitutive tensor are not negligible. In such a case, $D_a=\varepsilon^b_a E_b+\gamma^b_{a}H_b$, and $B_a=\mu^b_a H_b-\gamma^b_a E_b$ with $\gamma_{ab}$ the antisymmetric part of the constitutive tensor. In this case, the same derivation still stands, with the only difference that the antisymmetric parts of the constitutive tensor for the effective medium are given by $$\tilde{\gamma}^{\alpha\beta}=-\delta^{\alpha\beta\gamma}\frac{g_{0\gamma}}{g_{00}}-\sqrt{-g}\frac{g^{\alpha\delta}}{g_{00}}\gamma^{\beta}_\delta. $$ 
}}.

Non-linear media, with a Kerr-type non-linearity, can be treated analogously by promoting the dielectric and permeability tensors to explicitly depend on the field strengths. If also dispersion needs to be included in the game, we need to consider, as usual, the dispersion relation in frequency space in order to write it in a local form. Note that we can always write ${D}=\varepsilon_0{E}+{P}$ (and analogously for the magnetic field and excitation), moving all non-linearity and dispersion in the polarization (magnetization) vector. Thus, from eq.~\eqref{Maxelleff} we can conclude that the effective medium will give rise to an effective electric excitation
\begin{equation}
{D}_{eff}=\varepsilon_{\rm sp}(\varepsilon_0{E}+{P}),
\end{equation}
which can then be written, for the dispersive case of interest, locally in frequency space for the effective medium ``living'' in flat, Minkowski spacetime.

}

\section{Derivation of the NLSE: technical details}
Let us consider now Maxwell's equations for the effective medium, {thus in flat spacetime,} written in the usual notation
\begin{align}
    & \nabla\cdot B=0,\,\,\nabla\cdot D=0\\
    & \nabla\times E=-\partial_t B,\,\, \nabla\times H=\partial_t D,
\end{align}
with $D=\tilde{\varepsilon} E$ and $H=B/\tilde{\mu}$. Here, we consider the case of a spherically symmetric spacetime in isotropic coordinates. The metric can then be written, in full generality, as
\begin{equation}
    ds^2=-\left(\frac{B(t,r)}{A(t,r)}\right)^2 dt^2+a^2(t)A^4(t,r)\delta_{\alpha\beta}dx^{\alpha}dx^{\beta},
\end{equation}
with $r=\sqrt{\delta_{\alpha\beta}x^\alpha x^{\beta}}$, {$A(t,r),B(t,r)$ real functions, and $a(t)$ a scale factor analogous to the one appearing in FRLW spacetime}. 
Note that this metric can be rewritten as 
\begin{equation}\label{sphsimm}
    ds^2=\Omega^{-1}\left(-\frac{B^2(t,r)}{a^2(t)A^6(t,r)} dt^2+\delta_{\alpha\beta}dx^{\alpha}dx^{\beta}\right),
\end{equation}
where the ``conformal factor'' $\Omega=a(t)^{-2}A(t,r)^{-4}$. 

In particular, we specialize to the case for which $a(t)=1,\,A=A(r),\, B=B(r)$ and such that, {
in frequency space, $\tilde{\varepsilon}(E,r,\omega)=\varepsilon_0 A(r)^3B(r)^{-1}\left(1+\chi^{(1)}(\omega)+3\chi^{(3)}|E|^2/\Omega\right)$} 
and $\tilde{\mu}=\tilde{\mu}(r)= \mu_0{\mu_{\rm sp}=\mu_0} A(r)^3B(r)^{-1}$, 
i.e., we are considering a non-magnetic material, where all the magnetic properties are induced by the curved background, with a Kerr non-linearity. As we previously discussed, $\omega$ is the physical frequency defined with respect to the stationary observer $u^{\mu}$ that we assume to be the rest frame of the physical medium. {
The conformal factor $\Omega$ appearing in the non-linear term in $\tilde{\varepsilon}$ arises due to the fact that $E^aE_a$ in {curved} spacetime corresponds to $|E|^2/\Omega$, with $|E|^2=E^aE^b\delta_{ab}$ the flat spacetime norm squared of the electric strength, {in the flat spacetime of the effective medium, as can be easily seen directly from eq.~\eqref{sphsimm}.}}

{For the sake of notation clarity, let us emphasized that, in the following, tilded quantities refer to quantities pertaining to the effective medium in flat spacetime while the untilded ones represent the optical properties of the physical medium that is stationary in (physical) curved spacetime.}

In the following, we focus on Schwarzschild's spacetime, for which 
\begin{align}
   &A(r)=1+\frac{r_S}{4r}\\
   &B(r)=1-\frac{r_S}{4r},
\end{align}
where $r_S$ is the Schwarzschild's radius. 

From Maxwell's equations, taking the curl of the third one, we obtain
\begin{align}
    \nabla^2 E-\nabla\left(\nabla\cdot E\right)=\partial_t \left(\tilde{\mu}\partial_t D-B\times{\frac{\nabla\tilde{\mu}}{\tilde{\mu}}}\right).
\end{align}
{
and thus
\begin{align}\label{eq:wavenh_timedomain}
    \nabla^2 E-\nabla(\nabla\cdot E)-\tilde{\mu}\partial_t^2D=-(\nabla\log({\mu_{\rm sp}}))\times(\nabla\times E).
\end{align}
Note that this last expression is valid for {$\partial_t\tilde\mu=0$, which includes} the case of Schwarzschild spacetime. For a generic spherically symmetric metric, as in eq.~\eqref{sphsimm}, additional terms would be present due to the explicit time dependence of $\tilde{\mu}$. Moving now to frequency space, {where we indicate with $\nu$ the conjugate variable to the coordinate time $t$ in the flat spacetime of the effective medium,} and writing  $D=\tilde{\varepsilon}_\ell E+P_{\rm NL}$, where $\tilde{\varepsilon}_\ell$ is the linear part of the dielectric permeability and $P_{\rm NL}$ contains the nonlinear components of the polarization, we obtain 
\begin{align}\label{eq:wavenh_frequencydomain}
    \nabla^2 E-\nabla(\nabla\cdot E)+\tilde{\mu}\tilde{\varepsilon}_\ell \nu^2E=-\tilde{\mu} \nu^2 P_{\rm NL}-(\nabla\log({\mu_{\rm sp}}))\times(\nabla\times E).
\end{align}
This is our starting point for the derivation of the scalar NLSE. Note that, apart from the last term, the equation resembles the textbook wave equation modulo the inhomogeneity of the medium encoded in the coordinate dependence of $\tilde{\varepsilon}, \tilde{\mu}$~\cite{boyd2020nonlinear}.

Before starting the derivation of the NLSE, an observation is in order. In curved spacetime, the linear 
dispersion relation of the medium assumes the simple form, in the rest frame of the medium,
\begin{equation}
    n(\omega)= {c}\sqrt{\mu_0\varepsilon_0\varepsilon_r(\omega)}={c}\frac{\kappa}{\omega},
\end{equation}
with $\kappa$ the modulus of the spatial projection of the wave 4-vector {, $\varepsilon_r=1+\chi^{(1)}(\omega)$, and $\omega$ the physical frequency\footnote{This is connected to the frequency in flat spacetime  via $\omega=(\sqrt{-g_{00}})^{-1}\nu$.}, i.e., the frequency measured by an observer in curved spacetime}. 
Thus, {we write the dispersion relation for} our effective medium as 
\begin{equation}\label{ntilde13}
    \tilde{n}={c}\frac{\tilde{\kappa}}{\nu},
\end{equation}
where { $\tilde{n}=n_{\rm sp}n$ with $n_{\rm sp}^2=(\Omega |g_{00}|)^{-1}$}.
Eq.~\eqref{ntilde13} is the expression that we will use in deriving the NLSE. Note {once again} that here $\nu=\omega\sqrt{-g_{00}}$ where $\nu$ is the conjugate Fourier variable to the coordinate time $t$ in flat spacetime. {Since for consistency we want the two dispersion relations to be equivalent,}
we see that $\tilde{\kappa}=\kappa n_{\rm sp}\sqrt{-g_{00}}$. Once again, in the dispersive case, we will need to consider $\tilde{n}=\tilde{n}(\omega)$ since otherwise the two dispersion relations would not remain equivalent. }

\subsection{Derivation of the standard NLSE}
Before delving into the derivation of the NLSE for our effective, inhomogeneous medium, we summarize here the derivation of the NLSE in the standard case, following~\cite{boyd2020nonlinear}.

In the standard case of a homogeneous, non-magnetic material in flat spacetime, writing the displacement electric field $D$ as the sum of a linear part and the non-linear polarization, we have the wave equation 
{in frequency space
\begin{align}
    \nabla^2 E-\nabla(\nabla\cdot E)+\mu_0\varepsilon_\ell \nu^2E=-\mu_0\nu^2 P_{\rm NL}.
\end{align}
{Note that in this section we always work with untilded quantities that refer to the optical properties of the physical medium that is considered in flat spacetime. Indeed, in this case the effective medium  coincides with the physical one since the optical properties of flat spacetime are trivial. Note however that, as previously specified, from the next section we will go back to consider the case of curved spacetime. Thus, we will need to distinguish once again between physical and effective medium, with the latter represented by tilded quantities in flat spacetime.}

We recall that $\mu_0=1/(\varepsilon_0 c^2)$.
{We then neglect the vectorial operator $-\nabla(\nabla\cdot E)$ due to the fact that the homogeneous Maxwell equation for $D$ implies this term to be in general negligible} -- and get
\begin{align}
    \nabla^2 E(\nu)+\varepsilon_{r}(\nu) \frac{\nu^2}{c^2}E(\nu)=-\frac{\nu^2}{\varepsilon_0 c^2}P_{\rm NL}(\nu),
\end{align}
with {$\varepsilon_{r}(\nu)=\varepsilon_\ell/\varepsilon_0$} the linear, relative polarizability. 

{For a linearly polarized field, this equation becomes a scalar one. We can then} write the electric field as a slowly varying, {complex} amplitude $\mathcal{E}({\bf r},t)$ times a plane wave propagating in the $z$ direction with central frequency $\nu_0$
\begin{equation}\label{SVEAeq}
    E({\bf r},t)=\mathcal{E}({\bf r},t)e^{i(\kappa_0 z-\nu_0 t)}+cc.\text{,  where }\kappa_0=\frac{n(\nu_0)\nu_0}{c}.
\end{equation}
Using the Fourier transform w.r.t. $t$  for $E$ and the one for the amplitude\footnote{We follow~\cite{boyd2020nonlinear} in defining, 
\begin{align*}
    E(\mathbf{r},t)&=\int_{-\infty}^\infty \frac{d\nu}{2\pi} E(\mathbf{r},\nu)e^{-i\nu t}=\mathcal{E}(\mathbf{r},t)e^{i(\kappa_0 z-\nu_0 t)}+\mathcal{E}^*(\mathbf{r},t)e^{-i(\kappa_0 z-\nu_0 t)}\\ 
    &=\int_{-\infty}^\infty \frac{d\nu}{2\pi}\,\mathcal{E}(\mathbf{r},\nu)e^{-i(\nu+\nu_0) t}e^{i\kappa_0 z}+\int_{-\infty}^\infty \frac{d\nu}{2\pi}\,\mathcal{E}^*(\mathbf{r},\nu)e^{-i(\nu-\nu_0) t}e^{-i\kappa_0 z}\\
    &=\int_{-\infty}^\infty \frac{d\nu}{2\pi}\,\mathcal{E}(\mathbf{r},\nu-\nu_0)e^{-i\nu t}e^{i\kappa_0 z}+\int_{-\infty}^\infty \frac{d\nu}{2\pi}\,\mathcal{E}^*(\mathbf{r},\nu+\nu_0)e^{-i\nu t}e^{-i\kappa_0 z}.
\end{align*}
From these expressions we then obtain 
\begin{equation*}
    E(\mathbf{r},\nu)=\mathcal{E}(\mathbf{r},\nu-\nu_0)e^{i\kappa_0 z}+\mathcal{E}^*(\mathbf{r},\nu+\nu_0)e^{-i\kappa_0 z}
\end{equation*}} 
$\mathcal{E}$, {eq.~\eqref{SVEAeq}} can be rewritten in frequency space as a sum of terms dependent on $\nu\pm\nu_0$. We can then discard the {fast rotating, high frequency ($\nu+\nu_0$)} components. {Indeed, the slowly varying in time envelope $\mathcal{E}(\mathbf{r},t)$ in which we are interested does not possess high-frequency Fourier components~\cite{boyd2020nonlinear}. We thus obtain}   
\begin{equation}
    E({\bf r},\nu)\approx \mathcal{E}({\bf r},\nu-\nu_0)e^{i\kappa_0 z}.
\end{equation}
The scalar wave equation for the amplitude then becomes
\begin{align}
    \nabla^2_{\bot} \mathcal{E}+\partial^2_z \mathcal{E}+2i\kappa_0\partial_z \mathcal{E}+[\kappa^2(\nu)-\kappa_0^2]\mathcal{E}=-\frac{\nu^2}{\varepsilon_0 c^2}P_{\rm NL}e^{-i\kappa_0 z},
\end{align}
with $\kappa(\nu)={n(\nu)\nu}/{c}$. 
 At this point, we approximate $\kappa(\nu)$ as a power series in $\nu-\nu_0$
\begin{equation}
    \kappa(\nu)=\kappa_0+\kappa_1(\nu-\nu_0)+\mathscr{D},
\end{equation}
with $\mathscr{D}=\kappa_2 (\nu-\nu_0)^2/2+\mathcal{O}\left((\nu-\nu_0)^3\right)$, such that 
\begin{equation}
    \kappa(\nu)^2=\kappa_0^2+2\kappa_0\kappa_1(\nu-\nu_0)+2\kappa_0 \mathscr{D}+2\kappa_1 \mathscr{D} (\nu-\nu_0)+\kappa_1^2(\nu-\nu_0)^2+\mathscr{D}^2.
\end{equation}
Here $\kappa_1$ is the inverse of the group velocity $v_g$. We will neglect ${\mathscr{D}}^2$ terms and convert back to the time domain\footnote{This is achieved by multiplying the equation by $e^{-i(\nu-\nu_0)t}$ and integrating over all values of $\nu-\nu_0$. Recall that $P_{\rm NL}(\mathbf{r},t)=\int P_{\rm NL}(\mathbf{r},\nu)e^{-i\nu t}d\nu/2\pi$.} to obtain
\begin{align}
    \left(\nabla^2_{\bot}+\partial^2_z +2i\kappa_0(\partial_z+\kappa_1\partial_t)+2i\kappa_1 \bar{\mathscr{D}}\partial_t+2\kappa_0\bar{\mathscr{D}}-\kappa_1^2\partial^2_{t}\right)\mathcal{E}({\bf r},t) = \frac{1}{\varepsilon_0 c^2}\partial^2_{t}\left(P_{\rm NL}({\bf r},t)\right)e^{-i(\kappa_0 z-\nu_0 t)}.
\end{align}
Note that now $\bar{\mathscr{D}}$ is a differential operator with $\bar{\mathscr{D}}=-(\kappa_2/2) \partial^2_{t}+...$.
Finally, by writing also the polarization {$P_{\rm NL}({\bf r},t)=p({\bf r},t)e^{i\kappa_0 z-\nu_0 t}+c.c.$, i.e.,} as a {slowly-varying} amplitude $p({\bf r},t)$ times a plane wave $e^{i\kappa_0 z-\nu_0 t}$ propagating in the $z$ direction, one can see that the right-hand side (RHS) becomes\footnote{{
Here we can write 
\begin{align*}
 &\partial^2_{t}P({\bf r},t)e^{-i(\kappa_0 z-\nu_0 t)}= \partial^2_{t}\left(p({\bf r},t)e^{i(\kappa_0 z-\nu_0 t)}\right)e^{-i(\kappa_0 z-\nu_0 t)}\\
 &=\left(-\nu_0^2-2i\nu_0\partial_t+\partial^2_t\right)p({\bf r},t)
\end{align*}
}}
\begin{equation}
    \frac{1}{\varepsilon_0 c^2}\partial^2_{t}P_{\rm NL}({\bf r},t)e^{-i(\kappa_0 z-\nu_0 t)}=-\frac{\nu_0^2}{\varepsilon_0 c^2}\left(1+\frac{i}{\nu_0}\partial_t\right)^2 p({\bf r},t){+c.c.}.
\end{equation}

This is the starting point for implementing the slowly varying envelope approximation (SVEA). It usually involves moving to the frame moving with the pulse group velocity $\kappa_1^{-1}$, and then neglecting terms with second derivatives in the propagation direction. Let us sketch the procedure here:
\begin{itemize}
    \item The retarded frame is defined as $z'=z$ and $\tau=t-z/v_g=t-\kappa_1 z$.
    \item Thus, $\partial_z=\partial_{z'}-\kappa_1\partial_\tau$, and $\partial_t=\partial_\tau$ $\implies$ $\partial^2_z=\partial^2_{z'}-2\kappa_1\partial_{z'}\partial_\tau+\kappa_1^2\partial^2_{\tau}$.
    \item The wave equation thus becomes 
    \begin{align}
    \left(\nabla^2_{\bot}+\partial^2_{z'}\mathcal{E}-2\kappa_1\partial_{z'}\partial_{\tau}+2i\kappa_0\partial_{z'}+2i\kappa_1 \bar{\mathscr{D}}\partial_\tau+2\kappa_0\bar{\mathscr{D}}\right)\mathcal{E} = -\frac{\nu_0^2}{\varepsilon_0 c^2}\left(1+\frac{i}{\nu_0}\partial_\tau\right)^2 p.
    \end{align}
    \item Now the SVEA in space is valid when the pulse is longer than just a few wavelengths so that $\partial^2_{z'} \mathcal{E}\ll \kappa_0\partial_{z'} \mathcal{E}$. With this approximation
    \begin{align}\label{ot}
    \left(\nabla^2_{\bot}-2\kappa_1\partial_{z'}\partial_{\tau}+2i\kappa_0\partial_{z'}+2i\kappa_1 \bar{\mathscr{D}}\partial_\tau+2\kappa_0\bar{\mathscr{D}}\right)\mathcal{E} = -\frac{\nu_0^2}{\varepsilon_0 c^2}\left(1+\frac{i}{\nu_0}\partial_\tau\right)^2 p.
    \end{align}
    \item Moreover, one can also implement a SVEA in time since\footnote{This is not true, for example, in slow light materials.} {$\kappa_1/\kappa_0=(v_{\rm ph}/v_g)\nu_0^{-1}\approx\nu_0^{-1}$ where $v_{\rm ph}$ and $v_g$ are the phase and group velocities respectively}. When the pulse {length $T_{\rm pulse}$} is long enough to contain more than just a few optical cycles{, with $T_{\rm optical}=2\pi/\nu_0$}, within the envelope, then $(\kappa_1/\kappa_0) \partial_\tau\approx T_{\rm optical}/T_{\rm pulse}\ll 1$ so that 
    \begin{align}
    \left(\nabla^2_{\bot}+2i\kappa_0\partial_{z'}+2\kappa_0\bar{\mathscr{D}}\right)\mathcal{E}({\bf r},t) = -\frac{\nu_0^2}{\varepsilon_0 c^2} p({\bf r},t),
    \end{align}
    {where the time derivative of the slowly varying polarization envelope has been ignored, compared to the constant term, on the same basis that $T_{\rm optical}/T_{\rm pulse}\ll 1$. This approximation of the polarization term on the right hand side of eq.~\eqref{ot} is equivalent to neglecting the self-steepening effect~\cite{agrawal2000nonlinear}.}
\end{itemize}
}

\subsection{Setting up some important relations}
In the case of the inhomogeneous effective medium, we need to investigate some relation between the effective medium quantities and the one of the physical material before delving into the derivation of the NLSE. We have seen that the dielectric permeability and the magnetic one can be written in frequency space as
\begin{align}
    \tilde{\varepsilon}(E,r,\omega)&=\varepsilon_0\varepsilon_{\rm sp}(r)\left(1+\chi^{(1)}(\omega)+3\chi^{(3)} |E|^2/{\Omega}\right)\\
    \tilde{\mu}(r,\omega)&=\mu_{\rm sp}(r)\mu_0
\end{align}
where $\chi^{(1)}(\omega)$ is the material linear dielectric permeability, including the effect of dispersion. 
Note also that $\varepsilon_{\rm sp}(r)=\mu_{\rm sp}(r)$ in isotropic coordinates (that we are working with), so that $n_{\rm sp}(r)\equiv\sqrt{\varepsilon_{\rm sp}\mu_{\rm sp}}=\varepsilon_{\rm sp}(r)$. Thus, we have 
\begin{equation}
    \tilde{n}(\omega,r)=n_{\rm sp}(r)n(\omega)={c}\varepsilon_{\rm sp}(r)\sqrt{\mu_0\varepsilon_0(1+\chi^{(1)}(\omega))}.
\end{equation}

In the wave equation eq.~\eqref{eq:wavenh_frequencydomain}, we have the term $-\tilde{\mu}\tilde{\varepsilon}_\ell \nu^2E$ with $\tilde{\varepsilon}_\ell=\varepsilon_0\varepsilon_{\rm sp}(1+\chi_1(\omega))$ . In light of the previous considerations, this term can be written as 
\begin{equation}
    -\tilde{\mu}\tilde{\varepsilon}_\ell \nu^2 E=-(\tilde{n}^2/c^2) \nu^2 E.
\end{equation}
When we move to the frequency space for the effective medium, we use the conjugate variable ($\nu$) to Minkowski time. As we already noticed, this is related to the frequency measured by an observer at rest {with respect to the medium} in {curved spacetime} by $\nu=\omega\sqrt{-g_{00}}$. The effective dispersion relation is thus 
\begin{equation}
    \tilde{n}^2\nu^2={c^2}\tilde{\kappa}^2,
\end{equation}
as previously discussed (see eq.~\eqref{ntilde13}). 
In expanding in power series $\tilde{\kappa}$ around $\nu_0$ we will then have
\begin{equation}
   \tilde{\kappa}=\tilde{\kappa}_0+\tilde{\kappa}_1(\nu-\nu_0)+\tilde{\mathscr{D}}.
\end{equation}

By comparing the dispersion relation in curved spacetime and the one of the effective medium it is easy to see that 
\begin{align}\label{kappas2}
    &\tilde{\kappa}_0=\sqrt{-g_{00}}n_{\rm sp}\kappa_0\\
    &\tilde{\kappa}_1=n_{\rm sp}\kappa_1\label{kappas2_2}\\
    &\tilde{\kappa}_2=(n_{\rm sp}/\sqrt{-g_{00}})\kappa_2
    &\tilde{\mathscr{D}}=\frac{1}{2}\tilde{\kappa}_2(\nu-\nu_0)^2+...=n_{\rm sp}\sqrt{-g_{00}}{\mathscr{D}},
\end{align}
{where the $\kappa_i(\omega_0)$ appearing in these expressions are the analogues of their tilded versions, i.e.,}
{
\begin{align}
    &{\kappa}_0={\kappa}|_{\omega_0}\\
    &{\kappa}_1=\partial_\omega{\kappa}|_{\omega_0}\\
    &{\kappa}_2=\partial^2_\omega{\kappa}|_{\omega_0},
\end{align}
and refer to the tabulated optical properties of the physical medium we are considering.
}

The expression in eq.~\eqref{kappas2_2} implies that the group velocity in the effective medium is related to the physical one in curved spacetime by
\begin{equation}\label{hint:propervelocity}
    \tilde{v}_{\rm g}=v_{\rm g}/n_{\rm sp}.
\end{equation}
{Note that this is consistent with the way the phase-velocity in the effective medium is related to the one in curved spacetime via $\tilde{v}_{\rm ph}\equiv 1/\tilde{n}=1/(n_{\rm sp}n)=v_{\rm ph}/n_{\rm sp}$. More in general, this is consistent with the relation between the coordinate velocity $\tilde{v}=dx/dt$, characterizing the propagation in the effective medium in flat spacetime, and the {velocity with respect to an observer comoving with the dielectric} $v=d\chi/d\tau$ where $\tau=\sqrt{-g_{00}}t$ is the proper time with respect to the stationary observer and $\chi_i=x_i/\sqrt{\Omega}$ represents the proper length. Indeed, we see immediately that $v=d\chi/d\tau=\tilde{v}n_{\rm sp}$.
}

\subsection{Derivation of the NLSE for the effective medium}
First let us notice that, in order for the effective medium description to be equivalent to the physical one in curved spacetime, we need to require that:
\begin{enumerate}
    \item the dielectric permeability and magnetization are dependent on the radial coordinate with the expressions given in the previous section
    \item dispersion enters via the physical frequency $\omega=\nu/\sqrt{-g_{00}}$ which corresponds to a position dependent correction to the Fourier variable $\nu$.
\end{enumerate}
Note that the rest of the relations in the previous section are not necessary in the derivation of the NLSE, but they are nonetheless important for connecting the effective medium properties with the ones of the physical medium in curved spacetime.

In order to derive the NLSE in this case, we go back to the wave equation in eq.~\eqref{eq:wavenh_frequencydomain} that we report here for convenience
\begin{align}\label{eq:wavenh_frequencydomain1}
    \nabla^2 E-\nabla(\nabla\cdot E)+\tilde{\mu}\tilde{\varepsilon}_\ell \nu^2E=-\tilde{\mu} \nu^2 P_{\rm NL}-(\nabla\log({\mu_{\rm sp}}))\times(\nabla\times E).
\end{align}

To proceed further, as discussed in the main text, we can make use of the homogeneous Maxwell equation for $D$ in order to write

\begin{align*}
    \nabla\cdot D &= 0 \implies\nabla \cdot E  =   -  (\nabla \log\tilde\varepsilon_l)\cdot E - \frac{1}{\tilde\varepsilon_l}\nabla \cdot P_{\rm NL} \\
   - \nabla(\nabla\cdot E) &= (E\cdot\nabla)\nabla \log\tilde\varepsilon_l + \left((\nabla\log\tilde\varepsilon_l)\cdot \nabla\right)E + (\nabla\log\tilde\varepsilon_l)\times(\nabla\times E)+\nabla\left(\frac{1}{\tilde\varepsilon_l} (\nabla\cdot P_{\rm NL})\right),
\end{align*}
where we have used that $E\times \left(\nabla\times \nabla\log\tilde\varepsilon_l\right)=0$ since the curl of the gradient vanishes.
We obtain
\begin{align}\label{eq:wavenh_frequencydomain2}
    \nabla^2 E+\tilde{\mu}\tilde{\varepsilon}_\ell \nu^2E=-\tilde{\mu} \nu^2 P_{\rm NL}- \nabla\left(\frac{1}{\tilde\varepsilon_l} (\nabla\cdot P_{\rm NL})\right)-(E\cdot\nabla)\nabla \log\varepsilon_{\rm sp} - \left((\nabla\log\varepsilon_{\rm sp})\cdot \nabla\right)E-\left(\nabla\log\mu_{\rm sp}+\nabla\log\varepsilon_{\rm sp}\right)\times(\nabla\times E).
\end{align}

As discussed in the main text, eq.~\eqref{eq:wavenh_frequencydomain2} does not allow, in general, to write down a scalar propagation equation since even by starting from a linearly polarized electric field we end up having coupled equations between all the components of the electric field. This is in general also true whenever one does not ignore the vectorial term $\nabla(\nabla\cdot E)$. 

\begin{figure}[t!]
\includegraphics[scale=0.5]{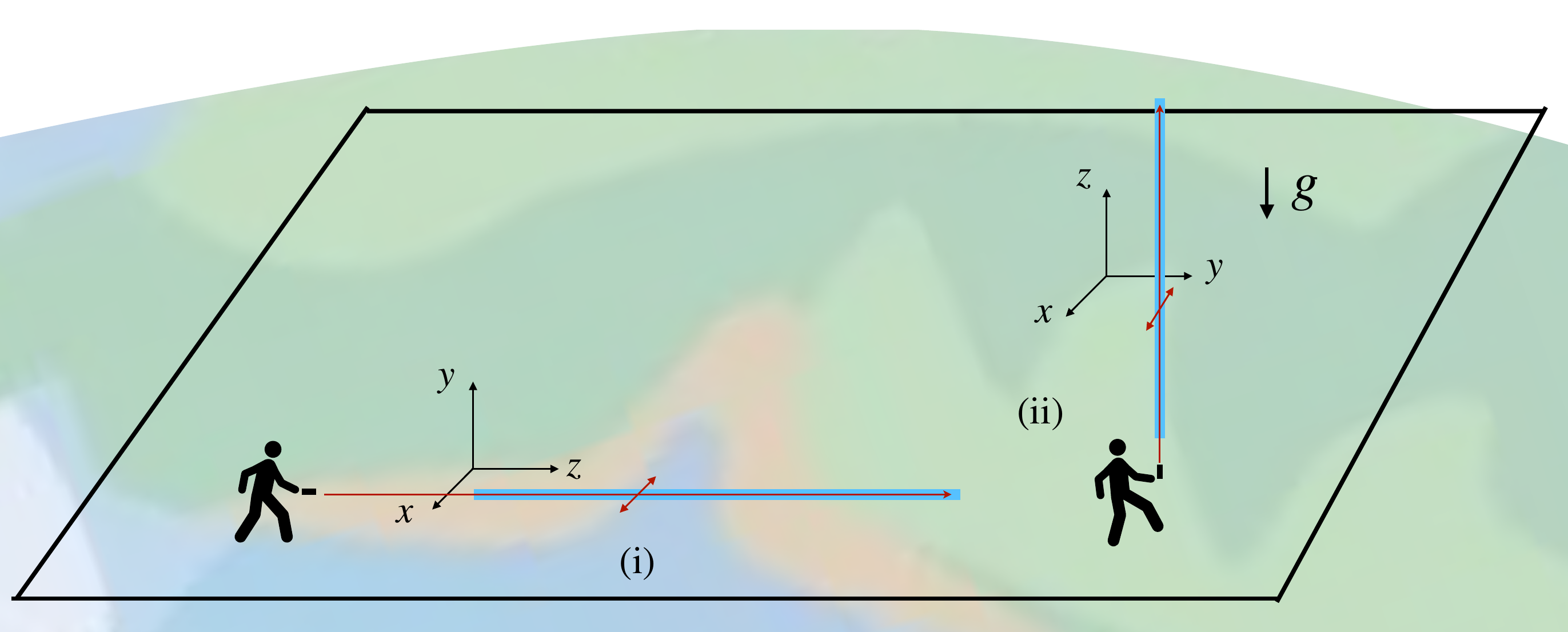}
\caption{Geometry of the problem. The two cases considered are labelled by (i) and (ii). In (i), the light pulse propagates in a horizontal fiber positioned at $r_\oplus=r\sim\text{constant}$. In (ii), the light pulse propagates in a vertically positioned fiber. }
\label{geometry}
\end{figure}

In order to bypass these problems, we resort to considering two cases of interest, which are the ones analyzed in the main text. See also Fig.~\ref{geometry}. Before doing so, let us emphasize that we will be interested in the specific case of a Kerr non-linear medium. Thus, we write the (slow envelope of the) non-linear polarization of the effective medium as
\begin{equation}
    p({\bf r},t)=3\varepsilon_0 n_{\rm sp}(r)\chi^{(3)}|\mathcal{E}|^2\mathcal{E}/{\Omega},
\end{equation}
which includes the non-linearity of the material and the contribution coming from the curved spacetime. 
Using the expression for $\kt_0$ in eq.~\eqref{kappas2}, the term containing the polarization can be written as 
\begin{align}\label{n2}
&-\frac{n_{\rm sp}(r)\nu_0^2}{2\kt_0\varepsilon_0{c^2}}p({\bf r},t)={-n_2\nu_0 n_{\rm sp}(r) \varepsilon_0 |\mathcal{E}|^2 \mathcal{E}/\Omega},
\end{align}
where {$n_2=3\chi^{(3)}/2n(\omega_0)c\varepsilon_0$} is the nonlinear index of the Kerr material. {As before, we are also going to neglect the self-steepening effect~\cite{agrawal2000nonlinear}.}
Furthermore, in our simulations we use the parameters of a 
single-mode, fused silica optical fiber employed in~\cite{philbin2008fiber} that we summarize here in Tab.~\ref{tab:parameters}.
\begin{table*}[h]
\centering
\begin{tabular}{llll}
    \hline
    \textbf{Symbol}  & & \textbf{Name} & \textbf{Value}  \\ \hline\hline
    Soliton pulse properties from \cite{philbin2008fiber}:\\ \hline
    $T_0$ & & Duration (this corresponds to 70\,fs total pulse length) & 40\,fs \\
    $E_s$ & & Generating pulse energy (not used, only for reference) & 5\,pJ \\
    $\lambda_0=2\pi c/\nu_0$ & & central soliton wavelength & 803\,nm \\
    \hline
    Fiber properties:\\ \hline
    $\kappa_0(\nu_0)$ & & $n(\nu_0) \nu_0 /c$, assuming $n(\nu_0)=1.5$ & $1.17 \cdot 10^7$\,/m \\
    $\kappa_1(\nu_0)$ & & $1/v_{\rm g}(\nu_0)$, assuming $v_{\rm g}(\nu_0)=0.65 c$ & $5.13176 \cdot 10^{-9}$\,s/m \\
    $\kappa_2(\nu_0)$ & & Group velocity dispersion from \cite{philbin2008fiber} & $-9.5 \cdot 10^{-27}$\,s$^2$/m \\
    $n_2$ & & Kerr non-linearity of silica from \cite{kabacinski2019nonlinear}
    &  $2.19 \cdot 10^{-20}$ m$^2$/W \\
    $A_{\rm eff}$ && Effective transverse mode area 
    & $\pi \left(1.6\,\mu\text{m}/2\right)^2$\\
    \hline 
    Properties of fused silica: \\ \hline
    $\mathcal{P}_{11\,22}$ & & Component for transverse stress of the photoelastic tensor from \cite{biegelsen1974photoelastic,primak1959photoelastic} & 0.271 \\
    $c_s$ & & Speed of sound tabulated in \cite{heraeusfusedsilica} & 5720\,m/s \\\hline
    Miscellaneous:\\ \hline
    $r_\oplus$&& Earth equatorial radius&6378137\,m\\
    $r_S$ (Earth) && Schwarzschild radius of Earth & $9 \cdot 10^{-3}$\,m\\
\end{tabular}
\caption{Specifics of all the parameters entering the numerical simulations of the NLSE(s). The material parameters are extrapolated from Philbin et al.~\cite{philbin2008fiber}, $\kappa_0/\nu_0 =1.5/c$ and $\kappa_1^{-1}\approx 0.65 c$. Consistently, we use the properties of the Crystal Fibre NL-PM 750 from NKT photonics~\cite{philbin2008fiber}.}
\label{tab:parameters}
\end{table*}

\subsubsection{Horizontal propagation}
As we have seen in the main text, considering linearly polarized light propagating -- in a medium stationary on Earth -- for distances much smaller than Earth's {radius, the horizontal motion can be considered as happening at constant radius  $r\geq r_{\oplus}$.}
We can then follow the derivation in \cite{philbin2008fiber} where the pulse propagation in a single-mode optical fiber was considered. 

In a nutshell, whenever the coefficients in eq.\eqref{eq:wavenh_frequencydomain1} are constant, so that the very last term vanishes {since $\nabla\log({\mu_{\rm sp}})=0$}, we find an equation 
{
\begin{align}
    \nabla^2 E-\nabla(\nabla\cdot E)+\tilde{\mu}\tilde{\varepsilon}_\ell \nu^2E=-\tilde{\mu} \nu^2 P_{\rm NL},
\end{align}
which is equivalent to eq.~(S1) of~\cite{philbin2008fiber} in frequency space.
}

{Following~\cite{philbin2008fiber}, and considering the propagation of light pulses in an optical fiber}, this equation can be solved by separation of variables between an amplitude that depends on the propagation direction and a vectorial part depending on the transverse directions{, i.e., $E(\nu,{\bf r})=E(\nu,z)U(\nu,x,y)$ in frequency space and with $U(\nu,x,y)$ a 3-dimensional vector}. By solving the eigenvalue problem for the transverse part{, we then remain with a one-dimensional problem given by}
\begin{equation}
    \partial_z^2 E(z)+\frac{\tilde{n}^2}{c^2}\nu^2 E(z)=-\tilde{\mu} \nu^2 P_{\rm NL}(z)
\end{equation}
where the refractive index {is set by the eigenvalue of the transverse fiber mode and} accounts for the property of the fiber's core and of the transverse profile. In our case, we can then assume to start directly from this equation, where the property of the effective medium accounts also for the non-trivial spacetime background via $n_{\rm sp}$.

At this point, the derivation of the NLSE proceeds as in the standard case discussed above. We introduce the field scalar amplitude via {$E(z,\nu)\approx \mathcal{E}(z,\nu-\nu_0)e^{i\kt_0 z}$ in our equation to obtain
\begin{align}
    \partial^2_z \mathcal{E}+2i\kt_0\partial_z \mathcal{E}+[\kt^2(\nu)-\kt_0^2]\mathcal{E}=-\nu^2\tilde{\mu}P_{\rm NL}e^{-i\kt_0 z}.
\end{align}
We then proceed as before by expanding 
\begin{equation}
   \tilde{\kappa}=\tilde{\kappa}_0+\tilde{\kappa}_1(\nu-\nu_0)+\tilde{\mathscr{D}}
\end{equation}
to get, neglecting ${\tilde{\mathscr{D}}}^2$ terms and converting back to the time domain,
\begin{align}
    \left(\partial^2_z +2i\kt_0(\partial_z+\kt_1\partial_t)+2i\kt_1 \bar{\tilde{\mathscr{D}}}\partial_t+2\kt_0 \bar{\tilde{\mathscr{D}}}-\kt_1^2\partial^2_{t}\right)\mathcal{E}(z,t) = \tilde{\mu}\partial^2_{t}\left(P_{\rm NL}(z,t)\right)e^{-i(\kt_0 z-\nu_0 t)},
\end{align}
where  $\bar{\tilde{\mathscr{D}}}=-(\kt_2/2) \partial^2_{t}+...$ in complete analogy with the standard derivation outlined above.

At this point, by neglecting the second derivatives in $z$ as well as terms $(\kt_1/\kt_0)\partial_t$ and using eq.~\eqref{n2} we arrive at eq.~\eqref{eq:soliton_diffeq_hor1} of the main text, i.e.,
\begin{align}
&i(\partial_z +\kt_1\partial_t)\mathcal{E}
-\frac{\kt_2}{2} \partial^2_t \mathcal{E} ={-n_2\nu_0 n_{\rm sp}(r_\oplus) \varepsilon_0 \frac{|\mathcal{E}|^2}{\Omega} \mathcal{E}}.
\end{align}
}

\subsubsection{Radial motion}
As we already discussed, in the case of vertical motion, in which we identify the radial direction with the propagation {direction along $z$ with $r=r_\oplus+z$}, the effective medium becomes a gradient-index medium with the refractive index changing along the propagation direction. We assume that all the quantities entering the wave equation depend solely on $z$. {Upon considering linearly polarized light along a direction orthogonal to $z$, we end up with the system of three decoupled equations in eq.~\eqref{eq:zdirection1} of the main text that we report here for completeness
\begin{align}
   &\partial_z^2E_{x(y)}+\tilde{\mu}\tilde{\varepsilon}_\ell\nu^2 E_{x(y)}=-\tilde{\mu}\nu^2P_{{\rm NL,} x(y)}-(\partial_z\ln\tilde{\varepsilon}_\ell)\partial_zE_{x(y)}+\left(\partial_z(\ln\tilde{\varepsilon}_\ell+\ln\tilde{\mu})\right)\partial_z E_{x(y)}\\  
   & \partial_z^2E_{z}+\tilde{\mu}\tilde{\varepsilon}_\ell\nu^2 E_{z}=-\tilde{\mu}\nu^2P_{{\rm NL,} z}-\partial_z\left(\frac{1}{\tilde{\varepsilon}_\ell}\partial_z P_{{\rm NL},z}\right)-2(\partial_z\ln\tilde{\varepsilon}_\ell)\partial_zE_{z}-E_z\partial_z^2\ln\tilde{\varepsilon}_\ell
\end{align}
}
We can then{: (1.) use the ansatz $E_x(z,t)\propto \mathcal{E}(z,t)e^{i(\tilde{\kappa}_0(z) z-\nu_0 t)}+cc.$; {(2.) proceed as before in expanding the dispersion relation around the central frequency, i.e., expanding $\kt(z,\nu)$ around $\nu_0$}; (3.) neglect $\tilde{\mathscr{D}}^2$ terms, to arrive at}
\begin{align}
&\frac{1}{2\kt_0}\partial_z^2\mathcal{E}+i(\partial_z +\kt_1\partial_t)\mathcal{E}-\frac{\kt_2}{2} \partial^2_t \mathcal{E} -2i \frac{\kt_1\kt_2}{4\kt_0}\partial_t^3 \mathcal{E}-\frac{\kt_1^2}{{2\kt_0}}\partial_t^2\mathcal{E}+2i\frac{\partial_z\kt_0}{2\kt_0} \mathcal{E}+ 2i z\frac{\partial_z\kt_0}{2\kt_0}\partial_z \mathcal{E}+i z\frac{\partial_z^2\kt_0}{{2\kt_0}} \mathcal{E}-z\partial_z\kt_0 \mathcal{E}-z^2\frac{(\partial_z\kt_0)^2}{2\kt_0} \mathcal{E} \\ \nonumber
&={-n_2\nu_0 n_{\rm sp}(r) \varepsilon_0 |\mathcal{E}|^2 \mathcal{E}/\Omega}+\frac{\partial_z\ln n_{\rm sp}}{2\kt_0}(i\kt_0\mathcal{E}+\partial_z \mathcal{E}+iz(\partial_z\kt_0)\,\,\mathcal{E}).
\end{align}
Upon using the SVEA approximation(s){, that entail that $\partial^2_{z} \mathcal{E}\ll \kappa_0\partial_{z} \mathcal{E}$ and $(\tilde{\kappa}_1/\tilde{\kappa}_0) \partial_t\ll 1$},
we then obtain the NLSE given by eq.~\eqref{eq:soliton_diffeq_vertical} in the main text. It should also be noted that, in the weak field approximation, the terms $-z^2\left({(\partial_z\kt_0)^2}/({2\kt_0})\right) \mathcal{E}$ and $({\partial_z\ln n_{\rm sp}})/({2\kt_0})iz(\partial_z\kt_0) \mathcal{E}$ are negligible since at least quadratic in $r_S/r_\oplus$. 

\section{Solution of the 1D equations}
\begin{figure}[t!]
\centering
\includegraphics[scale=0.5]{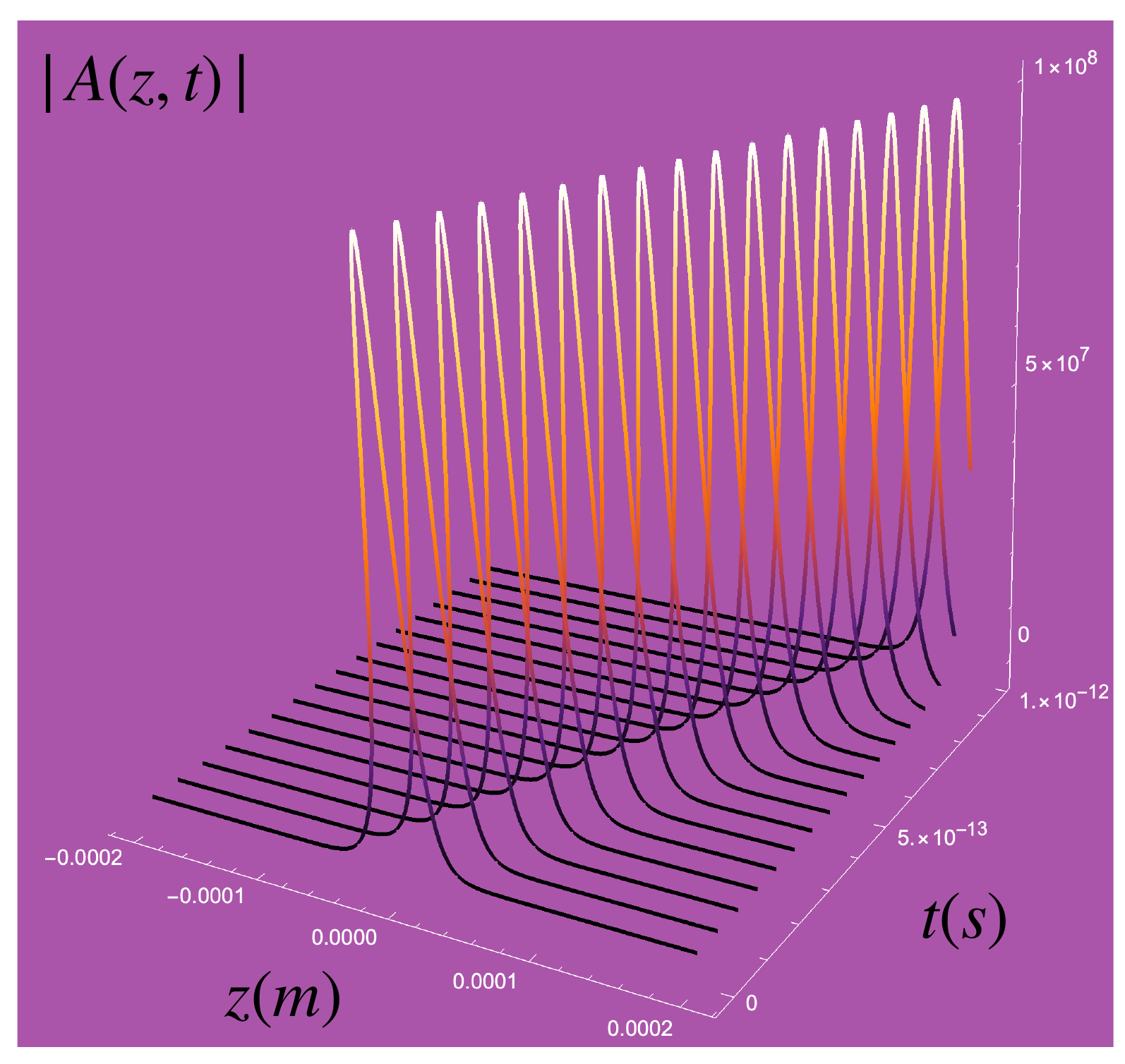}
\caption{Propagation of the 1D analytic soliton, eq.~\eqref{eq:soliton_diffeq_hor1}, for $r_S= 9\times 10^{-3}$\,m and $r_{\oplus}= 6\times 10^6$\,m.}
\label{geometry2}
\end{figure}

As discussed in the main text, in the case of horizontal propagation and considering a material with anomalous dispersion, we can solve eq.~\eqref{eq:soliton_diffeq_hor1} analytically. Borrowing the solution from {eq.(S74) of the supplementary material in~\cite{philbin2008fiber}} the analytical solution is given by (see also Fig.~\ref{geometry2})
\begin{equation}
\mathcal{E}(t,z)=\sqrt{\frac{\Omega | \kt_2| }{\nu_0 n_2 n_{\rm sp}\varepsilon_0 T_0^2 }}\cosh \left(\frac{t-\kt_1 z}{T_0}\right)^{-1}\exp \left(\frac{i z | \kt_2| }{2 T_0^2}\right),
\end{equation}
where $T_0$ is the pulse length, and $1/\kt_1$ is its speed of propagation. {This solution reduces exactly to eq.(S4) of~\cite{philbin2008fiber} in the limit of $r_S\to 0$.}
Note that the propagation speed of the soliton is $\tilde{v}_{\rm g}(\nu_0)=v_{\rm g}(\omega_0)/n_{\rm sp}$. This is exactly the proper velocity with respect to the observer's proper time and proper length in curved spacetime, as found above in eq. \eqref{hint:propervelocity}.

In the case of the vertical propagation, we solve eq.~\eqref{eq:soliton_diffeq_vertical} by way of the split-step Fourier method as showcased in~\cite{agrawal2000nonlinear}. In particular, we have adapted the Matlab code reported in~\cite{agrawal2000nonlinear} to our needs. In solving numerically the NLSE, we assign as initial temporal profile the soliton solution in flat spacetime of Philbin et al.~\cite{philbin2008fiber}, which coincides with the solution in the horizontal 1D propagation at $z=0$ and for $r_S\to 0$. 

Schematically, the split-step Fourier method consists in rewriting the NLSE as 
\begin{equation}
    \partial_z \mathcal{E}= \left(\hat{D}+\hat{N}\right)\mathcal{E},
\end{equation}
where the non-linear operator $\hat{N}=\hat{N}(z,|\mathcal{E}|^2)$ accounts for the non-linearity and the diffusive dynamics is enclosed in the operator $\hat{D}=\hat{D}(z,\partial_t)$. We then need to separate the action of the non-linear term and the dissipative one by dividing the propagation distance in small steps such that 
\begin{equation}
    \mathcal{E}(z+h,t)\approx e^{\hat{N}h/2}e^{\hat{D}h}e^{\hat{N}h/2} \mathcal{E}(z,t).
\end{equation}
This can be easily accomplished by alternating the use of the fast-Fourier/inverse Fourier transform algorithm in order to apply $\hat{D}$ in frequency space as a multiplicative operator and going back to the time domain at each step. Furthermore, since our operator $\hat{D}=\hat{D}(z,\partial_t)$ depends on the $z$ coordinate, a more precise implementation of the method would see to apply at each step $\exp\left(\int_{z}^{z+h}\hat{D}\right)$, which however is well approximated by $e^{\hat{D}h}$ in our simulations.

\section{Photoelasticity -- including the effect of material deformation on the refractive index}
As we have discussed so far, the optical medium in curved spacetime turns out to be equivalent to an effective one in flat spacetime, where the optical properties have a contribution coming from the curved spacetime background. However, whenever our physical medium is stationary in a curved spacetime{, i.e., it follows the trajectories of the timelike Killing vector}, it will also be subject to forces that can deform it. As discussed in the main text, deformations due to gravity of our physical medium lead to a change in the refractive index via the photoelastic effect~\cite{chen2006foundations}. 

Given our previous considerations, we will be interested in the effect of photoelasticity only for the vertical propagation equation. In order to include this effect and separate the contributions coming from the curvature of spacetime and the inertial acceleration of the fiber, we follow the discussion in~\cite{ratzel2018frequency} on the description of a deformable resonator.{ We choose to ignore the potential effects of photoelasticity on the nonlinear properties of the material, i.e., the nonlinear susceptibility $\chi^{(3)}$, as they would be mediated through different {mechanisms compared to the effect on the linear refractive index.}}

\begin{figure}[t!]
\centering
\includegraphics[scale=0.5]{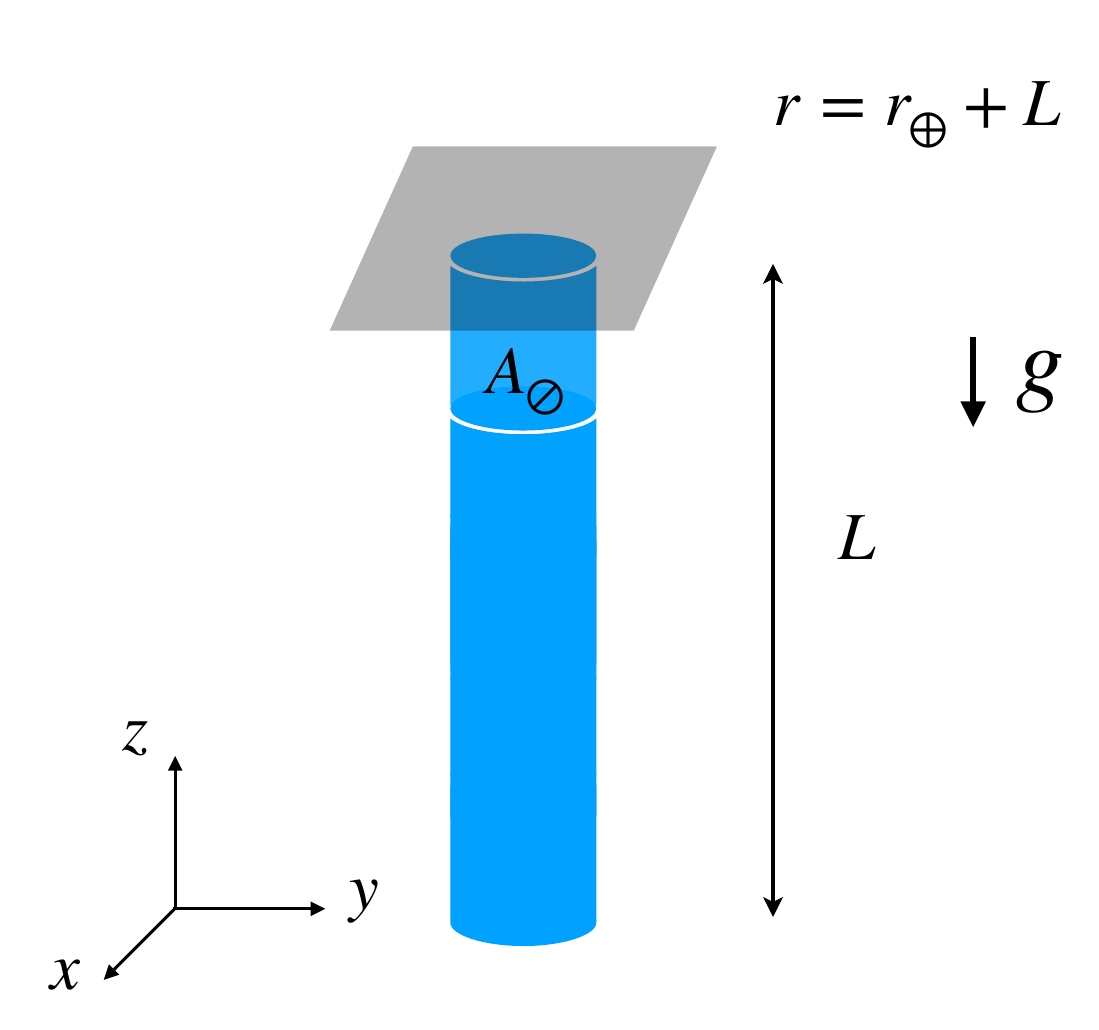}
\caption{Geometry of a fiber hanging in a weak gravitational field.}
\label{photoelastic}
\end{figure}

Consider then the situation depicted in Fig.~\ref{photoelastic}. A fiber of length $L$ and constant mass density {$\rho_m$} is hanging from a support located at $r=r_0\equiv r_\oplus+L$. In Schwarzschild spacetime, for an observer given by the stationary Killing vector $\partial_t/\|\partial_t\|$, the proper acceleration of the observer fixed at the support, {i.e., an observer at constant radius in isotropic coordinates,} and the local curvature projected into {the proper detector frame of this observer}
are given by~\cite{spengler2022influence}
\begin{align}
    {\bf{a}}^J &= \left(0,0, \frac{\frac{r_S}{2r_0^2}c^2}{\left(1-\frac{r_S}{4r_0}\right)\left(1+\frac{r_S}{4r_0}\right)^3} \right)\\
    R_{0J0J} &= \frac{r_S}{r_0^3\left(1+\frac{r_S}{4r_0}\right)^6}\left(\frac{1}{2},\frac{1}{2},-1\right),
\end{align}
where we chose for the $z$ direction to be aligned radially. {Furthermore, consistently with the notation we have used so far, we want to consider the origin of our coordinate at $r=r_\oplus$. This entails shifting $z\to z-L$ to translate the origin from the support at $r_0$ to $r=r_\oplus$.} {Note that the proper detector frame is determined by an orthonormal tetrad Fermi-Walker transported along the timelike trajectory of the support of the fiber which, in our set-up, corresponds to a stationary observer~\cite{misner1974gravitation}.}

{We can now compute the acceleration of test particles in the proper detector frame by following the derivation in~\cite{ratzel2018frequency}. At linear order in {$(z-L)/(r_\oplus+L)$}, the acceleration is given by {$a^z_p =-\left({\bf a}^z + c^2 R_{0z0z} (z-L)\right)$}. It should be noted that this expression is derived by neglecting acceleration squared terms in the proper detector frame metric as well as working at first order in the perturbations around flat spacetime (see discussion in~\cite{ratzel2018frequency}). This calls for care when wanting to extrapolate these expressions as generally valid.}
Each segment of the fiber is then stressed by the force {$F_z(z)$} of the parts of the fiber hanging ``below'' it
\begin{equation}
    {F_z(z) = \int_{0}^z {\mathrm d} z' \rho_m A_\oslash a^z_p(z')= -\rho_m A_\oslash c^2 \frac{r_S}{2r_0^2}\left( \frac{z}{\left(1-\frac{r_S}{4r_0}\right)\left(1+\frac{r_S}{4r_0}\right)^3} - \frac{z^2-2Lz}{r_0\left(1+\frac{r_S}{4r_0}\right)^6}\right),}
\end{equation}
where $A_\oslash$ is the cross-section of the fiber.

{More generally}, the fiber is subject to a stress $\sigma_{kl} = {F_k}/{A_l}$, where $F_k$ is the force in direction $\hat{e}_k$ and $A_l$ is the differential area normal to $\hat{e}_l$ upon which the force acts, caused by the inertial and tidal forces within the fiber. 
{As long as we are considering strains within the elastic limit of the material, which is the case of interest here,} we can employ Hooke's law and find that the strain in the fiber is $\mathcal{S}_{kl} =\sigma_{kl}/Y$, where $Y$ is the Young modulus of the material. The relation to the electric permeability tensor $\bm{\varepsilon}_r$ is then given by $\Delta(\bm{\varepsilon}_r)^{-1}_{kl} = \mathcal{P}_{kl\,mn}\mathcal{S}_{mn}$, where $\mathcal{P}$ is the photoelastic tensor~\cite{chen2006foundations}. The fact that the change in the inverse of $\varepsilon_r$ is linear in the strain holds for small or moderate strain. Limiting ourselves to isotropic materials, and a diagonal stress tensor, the equations reduce in complexity to 
\begin{align}
    \Delta(\bm{\varepsilon}_r)^{-1}_{kk} & = \mathcal{P}_{kk\,ll}\mathcal{S}_{ll}=\frac{\mathcal{P}_{kk\,ll}}{Y}\sigma_{ll}
\end{align}

In our set-up, the stress and then the strain on the fiber are given explicitly by
\begin{align}
    \sigma_{zz}(z) &= \frac{F(z)}{A_\oslash} = \rho_m c^2 \frac{r_S}{2r_0^2}\left( \frac{z}{\left(1-\frac{r_S}{4r_0}\right)\left(1+\frac{r_S}{4r_0}\right)^3} - \frac{z^2-2Lz}{r_0\left(1+\frac{r_S}{4r_0}\right)^6}\right)\\
    \mathcal{S}_{zz}(z) &= \frac{c^2}{c_s^2} \frac{r_S}{2r_0^2}\left( \frac{z}{\left(1-\frac{r_S}{4r_0}\right)\left(1+\frac{r_S}{4r_0}\right)^3} - \frac{z^2-2Lz}{r_0\left(1+\frac{r_S}{4r_0}\right)^6}\right),
\end{align}
where we used that the speed of sound in the fiber is $c_s = \sqrt{Y/\rho_m}$. {Note that the strain and stress have a positive sign due to the force being directed in the negative $z$ direction or, in other words, since we are considering an elongation of the fiber}. 
Due to the {axial symmetry of the problem,} 
 and {the irrelevance of two directions orthogonal to the $z$-axis for the 1D case}, the photoelastic tensor is a scalar.

The perturbation to the electric permeability, {promoting $\varepsilon_r\to\varepsilon_r+\Delta\varepsilon_r$}, is then also a scalar, and is given by
\begin{equation}\label{fullapprox}
    \Delta \varepsilon_r=-\frac{\varepsilon_r^2\Delta (\varepsilon_r^{-1})}{1+\varepsilon^0_r\Delta (\varepsilon_r^{-1})}\approx-(\varepsilon_r^0)^2\Delta (\varepsilon_r^{-1}),
\end{equation}
{where $\varepsilon_r$ indicates the electric permeability in the absence of photoelasticity} and the last expression holds whenever the photoelastic effect is a small correction to the material properties giving\footnote{{Note that here we have $\Delta (\varepsilon_r^{-1})=\mathcal{P}_{11\,22} \mathcal{S}_{zz}$. The indices are determined by the fact that we are considering an electric field linearly polarized in the $x$ direction, we identify $\{x,y,z\}\leftrightarrow \{1,2,3\}$, and we consider an isotropic material. Thus, (1) the only component of the perturbation tensor of interest is $\Delta (\varepsilon_r^{-1})_{11}$, (2) the only component of the strain is $S_{33}$, and (3) we have $\mathcal{P}_{11\,33}=\mathcal{P}_{11\,22}$. See Appendix D of~\cite{chen2006foundations} where the notation and the example of isotropic materials are discussed in detailed.}}

{\begin{equation}\label{appr}
    \Delta \varepsilon_r \approx -(\varepsilon_r^0)^2 \mathcal{P}_{11\,22} \mathcal{S}_{zz}(z) = - (\varepsilon_r^0)^2 \mathcal{P}_{11\,22} \frac{c^2}{c_s^2} \frac{r_S}{2r_0^2}\left( \frac{z}{\left(1-\frac{r_S}{4r_0}\right)\left(1+\frac{r_S}{4r_0}\right)^3} - \frac{z^2-2Lz}{r_0\left(1+\frac{r_S}{4r_0}\right)^6}\right).
\end{equation}}
\begin{figure}
\centering
\begin{subfigure}{.4\textwidth}
  \centering
  \includegraphics[width=1\linewidth]{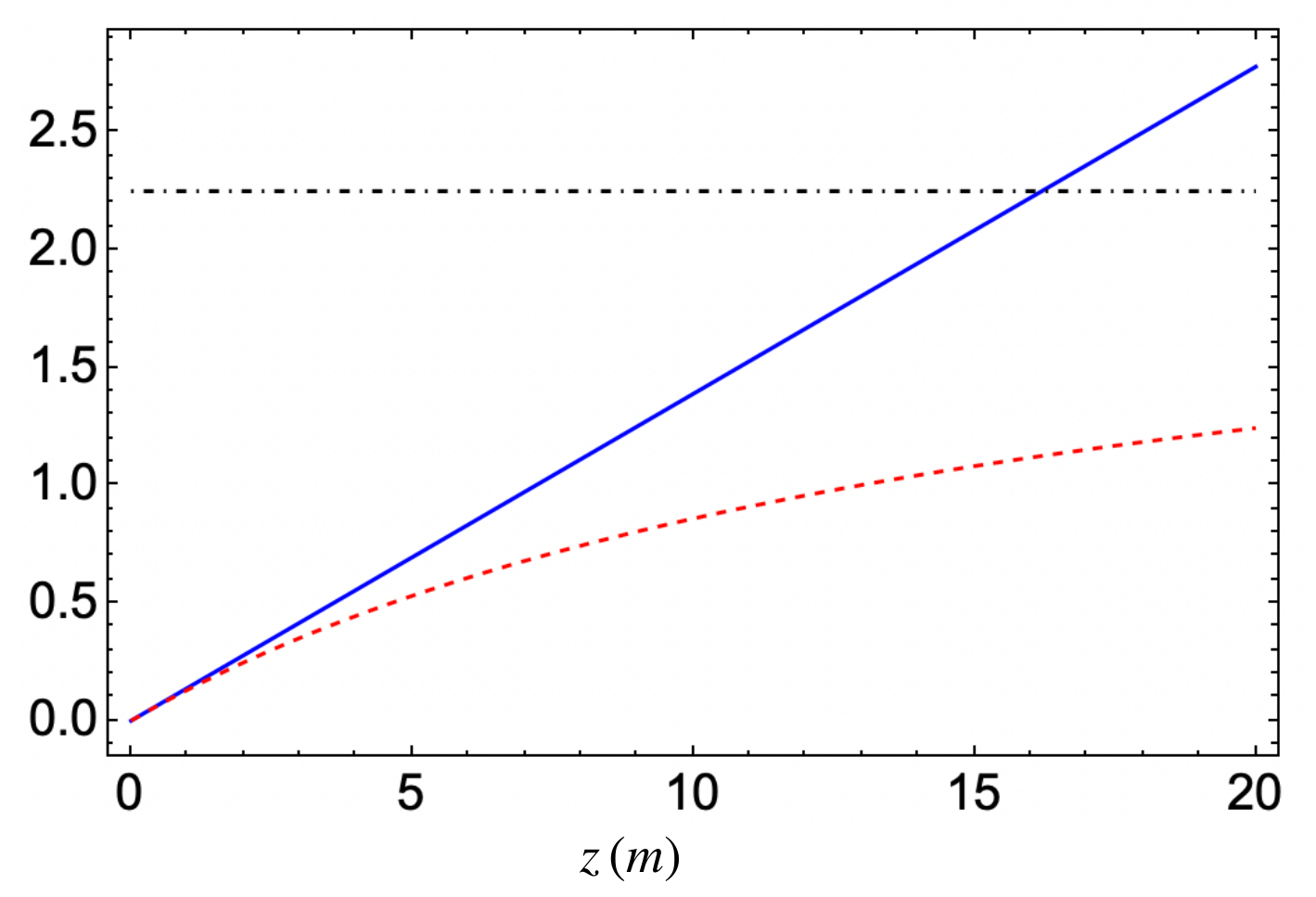}
  \label{fig:sub1}
\end{subfigure}%
\begin{subfigure}{.6\textwidth}
  \centering
  \includegraphics[width=1.1\linewidth]{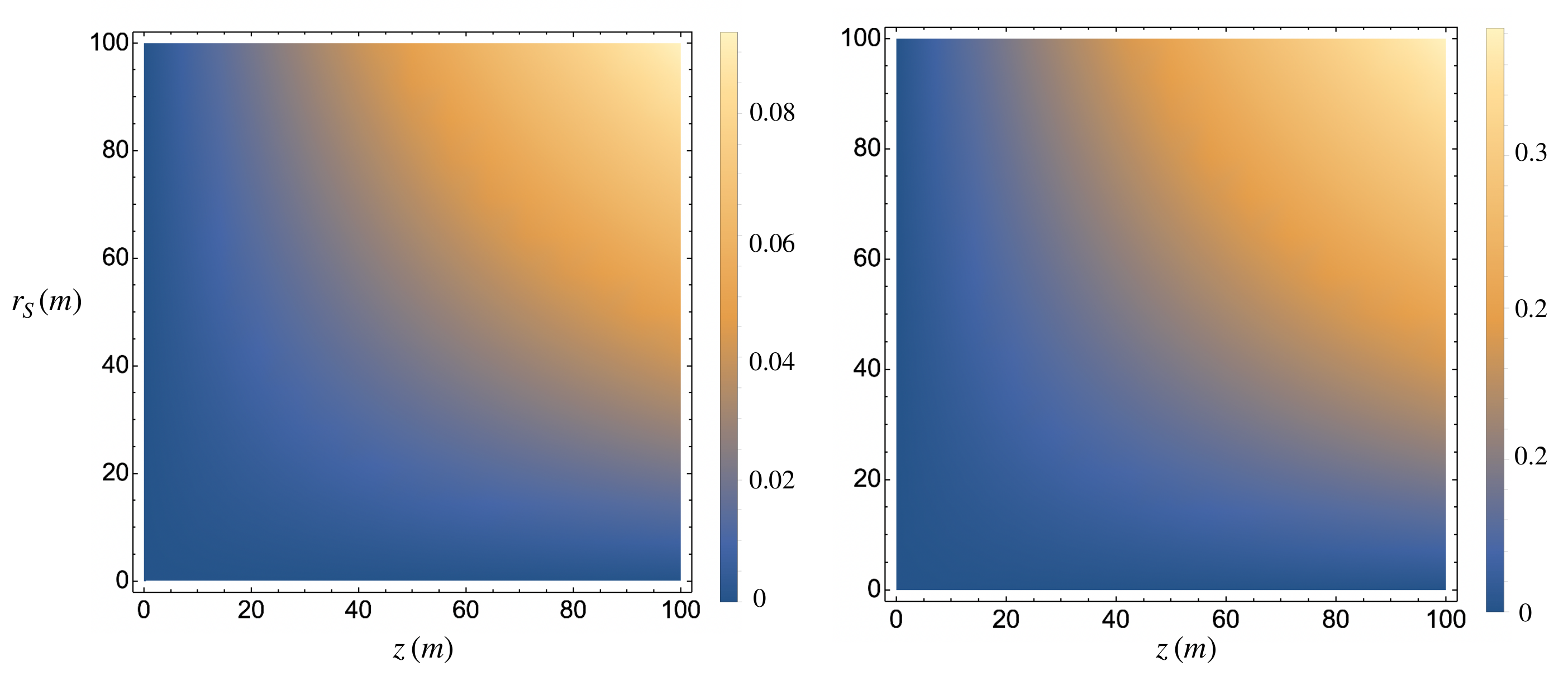}
  \label{fig:sub2}
\end{subfigure}
\caption{{Comparison between the full expression for $\Delta \varepsilon_r$ and the approximate one that are appearing in eq.~\eqref{fullapprox}. \textbf{Left panel:} Here we have used the parameters tabulated in Tab.~\ref{tab:parameters} and chosen a quite large $r_S=3$~km. {The solid, blue curve represents the approximate expression for $\Delta\varepsilon_r$, the dashed red curve the exact value of $\Delta\varepsilon_r$, while the dot-dashed black curve is the value of $\varepsilon_r$ in the absence of photoelasticity.} We see that (i) the full and approximate expressions start to deviate from propagation distances $\mathcal{O}(1{\rm m})$ onward and (ii) for relatively small propagation distances $\Delta\varepsilon_r$ is not anymore a small correction to the relative permeability $\varepsilon_r$ but becomes equal or greater than $\varepsilon_r$.} \textbf{Central panel:} Fractional difference between the full and approximated expressions for the photoelastic correction $\Delta\varepsilon_r$ i.e., ${\left(|\Delta\varepsilon_r|_{\rm approx}-|\Delta\varepsilon_r|_{\rm full}\right)}/{\left(|\Delta\varepsilon_r|_{\rm full}+|\Delta\varepsilon_r|_{\rm approx}\right)}$. Here $r_S$ goes from zero to $10^4$ times the Schwarzschild radius of Earth and the propagation distance reaches 100~m. We see that the difference between the two expressions remains below 10\%. \textbf{Right panel:} $|\Delta\varepsilon_r|_{\rm full}$. The value of $\Delta\varepsilon_r$, for $r_S$ from zero to $10^4$ times the Schwarzschild radius of Earth and propagation distance up to 100~m, is always well below the value of the relative permeability $\varepsilon_r\approx 2.25$.  }
\label{figdeltae}
\end{figure}

{Photoelasticity represents an additional correction to the electric permeability on top of the other effects accounting for the effective medium as described in the previous sections. For fused silica, the tabulated values in~\cite{biegelsen1974photoelastic,primak1959photoelastic} give $\mathcal{P}_{11\,22} = 0.271$ and $c_s= 5720$\,m/s~\cite{heraeusfusedsilica}. Then, from eq.~\eqref{fullapprox}, for a 10\,cm long fiber in the gravitational field of Earth, the contribution of the inertial acceleration {(first term in eq.~\eqref{appr})} at the end of the fiber to $\Delta\varepsilon_r$ is on the order $10^{-8}$ while the tidal acceleration {(second term in eq.~\eqref{appr})} contributes a term of order $10^{-16}$. Note that, while the tidal contribution is clearly negligible, the correction to the relative permeability induced by the inertial acceleration is between one and two orders of magnitude greater than the correction due to the vacuum curved spacetime optical properties in our effective picture {as quantified by $1-\varepsilon_{\rm np}\sim 10^{-9}$}.}

{It is easy to check that, considering a $\mathcal{P}_{11\,22} \approx 0.271$, the approximate expression in eq.~\eqref{appr} will start to fail around a propagation length of 2~m if we consider to be at one Earth's radius distance from an object whose mass corresponds to a Schwarzschild radius of 3~km. Indeed, Fig.~\ref{figdeltae} shows this failure as well as the fact that for such extreme values of $r_S$, $\Delta\varepsilon_r$ starts to be comparable or greater than $\varepsilon_r$ at propagation distances less than 10~m. At the same time, the same figure shows that, for $r_S$ up to $10^4$ times the one of Earth, both the conditions for the validity of the approximate expression for $\Delta\varepsilon_r$ and the fact that the correction to $\varepsilon_r$ is small are well satisfied. }

\section{{Conditions for validity of the unidirectional approach}}
As discussed in the main text, when considering the propagation of a light pulse in a gradient-index medium, we should account for the fact that the position-dependent refractive index will cause some light to be backscattered -- {this effect has also technological application in distributed acoustic sensing for seismology, see~\cite{williams2022listening} and references therein.} However, when solving the NLSE using the SSF method, the boundary condition completely ignores this fact -- it would require already knowing the solution to include the backscattered light in the boundary condition. That this is a drawback of using the NLSE -- {which is a unidirectional equation for the validity of which, by definition, back-propagating fields must be negligible --} in conjunction with the SSF method is well known~\cite{PhysRevLett.89.283902,PhysRevA.81.013819}. 

However, back-propagating fields
cannot always be simply ignored. A formalism fully accounting for this issue would require to solve a system of coupled bidirectional equations, or just solve the full Maxwell equations. However, as argued in~\cite{PhysRevA.81.013819}, we can define conditions that guarantee us that the backward reflected light is negligible. In our case, this sets a restriction on the parameter space that we can explore, where the description given by our solution to the NLSE can be trusted. Essentially, this regime corresponds to the one of weak-field and not large propagation distances. Indeed, physically, for vertical propagation, longer propagation distances and stronger gravitational accelerations would imply greater changes to the refractive index {giving potentially rise to non-negligible back-propagating fields.} 
To make this observation more quantitative, we follow here~\cite{PhysRevA.81.013819} where a more detailed discussion can be found.

We start from eq.~\eqref{eq:zdirection1} that we report here for convenience
\begin{equation}\label{starthere}
    \partial_z^2E_{x}+\tilde{\mu}\tilde{\varepsilon}_\ell\nu^2 E_{x}=-\tilde{\mu}\nu^2P_{{\rm NL,} x}+(\partial_z\ln\tilde{\mu})\partial_z E_{x}.
\end{equation}
Following~\cite{PhysRevA.81.013819}, we can rewrite this equation as 
\begin{equation}
   (\partial^2_z+\beta^2)E_x(z)=-\mathcal{Q}(z,E_x), 
\end{equation}
where $\beta$ is a reference momentum that can contain the dispersive character of the physical medium but no $z$ dependence and that forms our underlying dynamics on top of which we have some perturbation encoded in $\mathcal{Q}$, {\textit{the residual terms}}. In our case, a sensible choice for $\beta$ is 
\begin{equation}
    \beta^2=\frac{n_0^2\nu^2}{c^2},
\end{equation}
where $n_0= \sqrt{\varepsilon_r(\bar{\omega}_0)}$ is the material refractive index without any additional effect from the spacetime (and ignoring the effect of redshift {combined} with the dispersion of the material) {and not accounting for the photoelasticity}. With this choice we have\footnote{
{
In a nutshell, from eq.~\eqref{starthere} we have
\begin{align*}
    &\partial_z^2E_{x}+\mu_0\epsilon_0(\varepsilon_r+\Delta\varepsilon_r)\nu^2\mu_{sp}\varepsilon_{sp} E_{x}=-\mu_0\mu_{sp}\nu^2P_{{\rm NL,} x}+(\partial_z\ln\tilde{\mu})\partial_z E_{x}
\end{align*}
Writing then $\mu_0\epsilon_0(\varepsilon_r+\Delta\varepsilon_r)\nu^2\mu_{sp}\varepsilon_{sp}=\mu_0\epsilon_0\nu^2\varepsilon_r(1-(1-n_{sp}^2))+\mu_0\epsilon_0\nu^2n_{sp}^2\Delta\varepsilon_r$
we arrive at 
\begin{align*}
    &\partial_z^2E_{x}+\beta^2 E_x=\beta^2{\left[1-n_{\rm sp}^2\left(1+\frac{\Delta\varepsilon_r}{\varepsilon_r}\right)\right]}E_x
    -\mu_0\mu_{\rm sp}\nu^2 P_{{\rm NL,} x}+(\partial_z\ln\tilde{\mu})\partial_z E_{x}
\end{align*}
}
} 
\begin{equation}
    -\mathcal{Q}(z,E_x)=-\tilde{\mu}\nu^2P_{NL,x}(z)+(\partial_z\log\tilde{\mu})\partial_z E_x(z)+\beta^2[1-n_{\rm sp}^2(1+\Delta\varepsilon_r/\varepsilon_r)]E_x(z).
\end{equation}

Now, we can decompose the field in forward and backward directed {(in time)} fields $E_x=E_++E_-$ and find the equivalent system of two equations~\cite{PhysRevA.81.013819}
\begin{equation}\label{eqfirst}
     \partial_z E_{\pm}=\pm i\beta E_{\pm}\pm\frac{i}{2\beta}\mathcal{Q}.
\end{equation}
The question is then when, starting with $E_-=0$, $E_-$ remains negligible. Indeed, if $E_-$ remains negligible then we are left with a unidirectional equation and, more importantly, we know that the reflected light can be safely neglected even in comparison with the unperturbed propagation in flat spacetime in a linear medium. 

Photoelasticity is the main culprit for the possible significance of the reflected light since, as we have argued before, it is the dominant effect giving rise to an effective gradient-index medium. It enters only in the term linear in the electric field. Thus, we focus solely on this term in the following. As discussed in~\cite{PhysRevA.81.013819}, the first condition for the backward propagating light to be negligible is  that the residual terms contained in the term $\mathcal{Q}/2\beta$ {in eq.~\eqref{eqfirst}} are negligible with respect to $\beta E_x$. This translates to the condition
\begin{equation}\label{slow}
    {1-n_{\rm sp}^2(1+\Delta\varepsilon_r/\varepsilon_r)}\ll 1.
\end{equation}

The second condition arises from considering 
the backward-evolving part of $E_{-}${(cf. the discussion in the appendix of~\cite{PhysRevA.81.013819})} 
\begin{equation}
    \partial_z E_{-,\,\rm backwards}\approx\frac{\partial_z \chi}{(k+\beta)^2},
\end{equation}
where $\mathcal{Q}=\chi E_x$ and $k^2(z)\equiv\beta^2+\frac{\mathcal{Q}(z)}{E_x(z)}$. {For small $\mathcal{Q}$ and ignoring} non-linearities, {requiring that the change in the medium parameters does not cause significant back-propagation on the order of a wavelength leads to
\begin{equation}\label{noacc}
    \frac{\partial_z \left(n_{\rm sp}^2(1+\Delta\varepsilon_r/\varepsilon_r)\right)}{(3/2+n_{\rm sp}^2(1+\Delta\varepsilon_r/\varepsilon_r))^2}\ll \beta.
\end{equation}
}
This \textit{no-accumulation} condition 
requires that the derivative of the backpropagating fields is negligible and encodes the fact that there is no-accumulation of the reflected light giving in the end a non-negligible contribution. 

\begin{figure}[t!]
\centering
\includegraphics[scale=0.6]{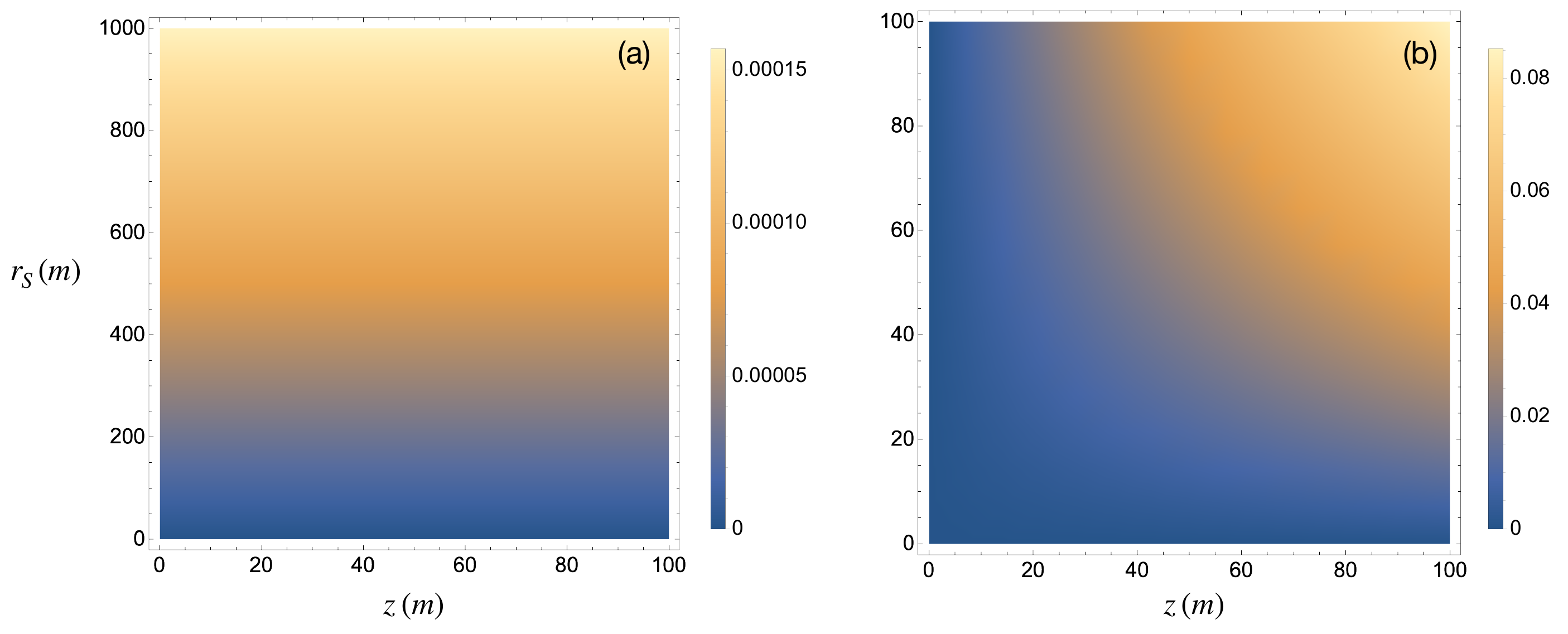}
\caption{\textbf{Panel (a)} shows the right-hand side of the slow-evolution condition of eq.~\eqref{slow} in the case in which photoelasticity is not considered; \textbf{Panel (b)} shows the same when also photoelasticity in included. We see that without photoelasticity the condition is very well satisfied for a large set of parameters. When including photoelasticity, we see violations of the condition for values of $r_S$ or the propagation length for which, from {Fig.~\ref{figdeltae} and the corresponding discussion,} we know that the $\Delta\varepsilon_r$ starts to be not anymore a small correction to the relative permeability. }
\label{figslow}
\end{figure}

\begin{figure}[t!]
\centering
\includegraphics[scale=0.6]{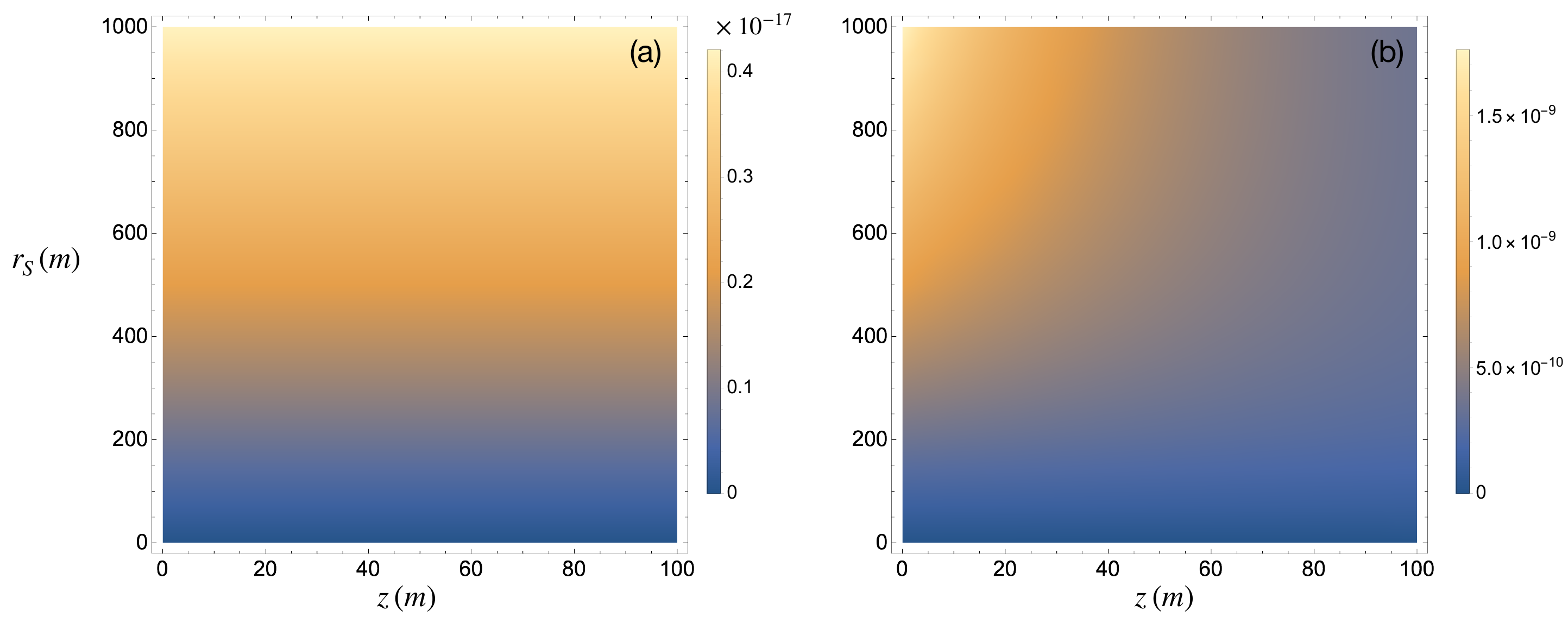}
\caption{\textbf{Panel (a)} shows the no-accumulation condition {by depicting the ratio between the left and the right-hand sides of eq.~\eqref{noacc}} in the case in which photoelasticity is not considered; \textbf{Panel (b)} shows the same when also photoelasticity in included. We see that this condition is actually well satisfied in both cases, while it remains that without photoelasticity the condition is much better satisfied. This analysis shows that the slow-evolution condition is the relevant one for the problem that we are considering.}
\label{figNoaccu}
\end{figure}

\begin{figure}
    \centering
\includegraphics[scale=0.5]{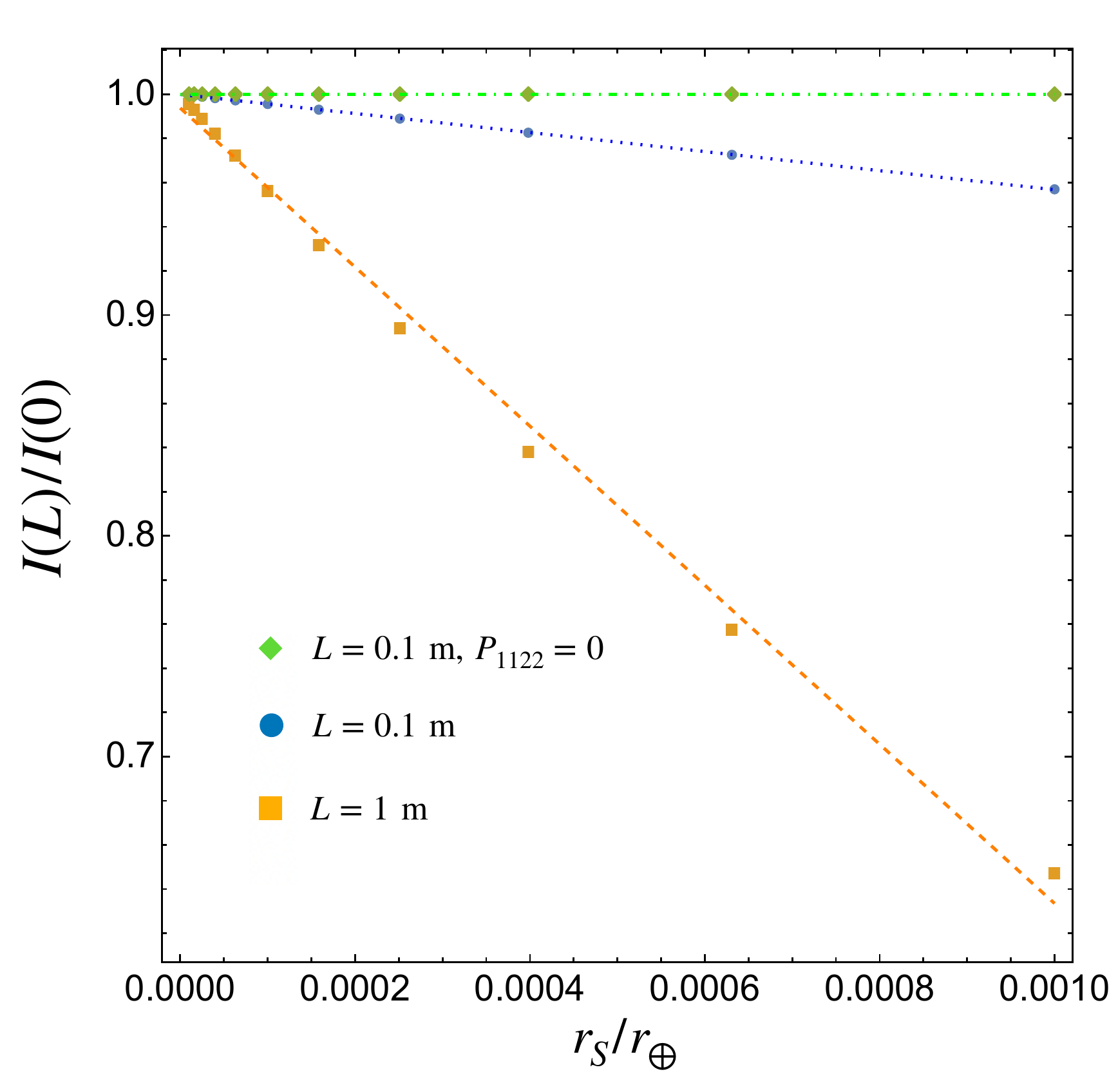}
    \caption{Energy loss due to photoelasticity. We show the ratio between the final and initial energy $I(L)/I(0)\sim\left.\varepsilon_r\int dt |E|^2\right|_{z=L}/\left.\varepsilon_r\int dt |E|^2\right|_{z=0}$, in the proper detector frame. The orange, square points correspond to the case of a propagation length of 1\,m while the blue, round points to a propagation length of $0.1$~m. The lines represent the linear fit of the corresponding data. We obtain slopes of $-360.3 r_s/r_\oplus$ and $- 43.2 r_s/r_\oplus$ respectively. The green, rhomboidal points correspond to the case without photoelasticity and are compatible with energy conservation {up to a negligible energy loss accounted for by purely gravitational redshift}.}
    \label{fig:Energyloss}
\end{figure}

{From Fig.~\ref{figslow} and Fig.~\ref{figNoaccu} we can see that,}
 in the absence of photoelasticity, i.e., considering a rigid dielectric, these conditions are very well satisfied for the parameters in our simulations also when considering relatively large values of $r_S$ and propagation lengths. When turning on photoelasticity, the situation changes, and we can arrive to regimes of large $r_S$ and large propagation distances where the conditions are not satisfied anymore. In particular, from Fig.~\ref{figslow} and Fig.~\ref{figNoaccu} we see that the main limiting factor is the slow-evolution condition. 
However, it should be noted that the slow-evolution condition starts to be violated in the same range of parameters in which $\Delta\varepsilon_r$ cannot anymore be considered a small correction and when it is arguable if the treatment of the photoelasticity as linear in the stresses is valid. To corroborate these observations, in the regime in which the slow-evolution condition is clearly violated we observe a non-negligible energy loss in the numerical solutions of the vertical propagation equation (see Fig.~\ref{fig:Energyloss}).

{
\section{Coefficients for the numerical simulations:}
Finally, we report here the explicit expressions for the different coefficients entering the vertical propagation equation that we use in our simulations.  

First, let us recall that, when including photoelasticity, we have
\begin{equation}
    n(\omega)=\sqrt{\varepsilon_r(\omega)+\Delta \varepsilon_r(\omega)}=\sqrt{1+\chi_1(\omega)+\Delta \varepsilon_r(\omega)}={\sqrt{n_0(\omega)^2+\Delta \varepsilon_r(\omega)}},
\end{equation}
{where $n_0(\omega)$ is the refractive index in the absence of photoelasticity.}
{We can then proceed to compute all the $\kappa_i$ coefficients of interest
\begin{align}
    &{\kappa}_0={\kappa}|_{\omega_0}\\
    &{\kappa}_1=\partial_\omega{\kappa}|_{\omega_0}\\
    &{\kappa}_2=\partial^2_\omega{\kappa}|_{\omega_0},
\end{align}
where $\kappa=n\omega/c$.
}
{We start from $\kappa_0$, where we have}
\begin{align}
    &\kappa_0(\omega_0)=\frac{n(\omega_0)\omega_0}{c}.
\end{align}
{We now consider the case in which the pulse propagates from the bottom of the vertically oriented fiber, which is the case we simulate numerically. We thus refer the various quantities of interest to the initial physical frequency $\bar{\omega}_0$, i.e., the frequency measured by the stationary observer at the bottom of the fiber. We can then write}
\begin{align}\label{eq:k0}
    &\kappa_0(\omega_0)=\frac{n\left(\bar{\omega}_0\frac{\sqrt{-g_{00}(r_{\oplus})}}{\sqrt{-g_{00}(r_{\oplus}+z)}}\right)\bar{\omega}_0\frac{\sqrt{-g_{00}(r_{\oplus})}}{\sqrt{-g_{00}(r_{\oplus}+z)}}}{c}.
\end{align}
Note that, even for extreme values of $r_S$ and $z$, like $r_S=10^{-2}r_\oplus$ and $z=100$~m we have that ${1-\sqrt{-g_{00}(r_{\oplus})}}/{\sqrt{-g_{00}(r_{\oplus}+z)}}$ is negligible when considering the dispersive properties of realistic materials at the optical frequencies of interest, i.e., the changes would be on scales way too fine-grained with respect to the tabulated values of the refractive index at the $\mu$m scale~\cite{kitamura2007optical}.
To account for this fact, calling $\zeta=1-\sqrt{-g_{00}(r_\oplus)}/\sqrt{-g_{00}(r_\oplus+z)}$ 
we perform an expansion of eq.~\eqref{eq:k0} at the first order in $\zeta$. 

{In the following we report the expressions for all the coefficients necessary to simulate the vertical propagation of the pulse at first order in $\zeta$. Note however that, for what concerns the simulations reported in the work, we can always safely neglect also the corrections to the zeroth order terms for all the $\kt_i$. The same holds true also for the terms $\partial_z\kt_0$ and $\partial^2_z\kt_0$ as far as photoelasticity is considered since the $z-$dependence is dominated by the photoelasticity. However, when considering the case with no photoelasticity, neglecting the $z-$dependence coming from the redshift factors in $\kappa_0$ amounts to a relative error of one part in $10^3$. While still small, we have performed the simulations in which photoelasticity is not included considering also the $\zeta$ corrections in full to account for this small discrepancy.}

The full expression including all the corrections at order $\zeta$ are reported in the following. Starting with $\kappa_0$ we have
\begin{align}
    \kappa_0(\omega_0)&\approx {\kappa_0(\bar{\omega}_0)}-\zeta\bar{\omega}_0\kappa_1(\bar{\omega}_0)\\ \nonumber
    &={\kappa^{\rm b}_0\sqrt{1+\frac{\Delta\varepsilon_r}{\varepsilon_r}}}-\zeta\left(\kappa^{\rm b}_0\sqrt{1+\frac{\Delta\varepsilon_r}{\varepsilon_r}}+\frac{\bar{\omega}_0^2}{c}\frac{\varepsilon_r'+\Delta\varepsilon_r'}{2n_0\sqrt{1+\frac{\Delta\varepsilon_r}{\varepsilon_r}}}\right),
\end{align}
where $\kappa_1(\bar{\omega}_0)=(n(\bar{\omega}_0)+\bar{\omega}_0\partial_\omega n|_{\bar{\omega}_0})/c$, while $n_0$ and $\kappa^{\rm b}_0$ are the tabulated refractive index and corresponding $\kappa_0$ of the material, {without photoelasticity}, i.e. $n_0=\sqrt{\varepsilon_r(\bar{\omega}_0)}$ and $\kappa^{\rm b}_0=\bar{\omega}_0n_0/c$, and a prime indicates the derivative with respect to the frequency. 

Note that $ \varepsilon_r'=\chi_1'$. We derive the expression for the latter, in terms of tabulated values, below. 
Before doing so, however, let us compute the derivatives of $\kt_0$ at first order in $\zeta$. Using the fact that $\kt_0=\sqrt{-g_{00}(r_\oplus+z)}n_{\rm sp}(z)\kappa_0(\omega_0)$, we find
{\small
\begin{align}
    {\partial_z\kt_0\approx}&{\kappa^{\rm b}_0 \sqrt{-g_{00}(r_\oplus)} \left(\partial_z n_{\rm sp} \sqrt{\frac{\Delta\varepsilon_r}{\varepsilon_r}+1}+\frac{n_{\rm sp} \partial_z\Delta\varepsilon_r}{2 \varepsilon_r \sqrt{\frac{\Delta\varepsilon_r}{\varepsilon_r}+1}}\right)}+ \frac{\bar{\omega}_0^2}{4 c (\Delta \varepsilon_r+\varepsilon_r (\bar{\omega}_0))^{3/2}} \left[-2 \zeta (\partial_z n_{\rm sp}) \sqrt{-g_{00}\left(r_\oplus+z\right)} (\Delta \varepsilon_r+\varepsilon_r (\bar{\omega}_0)) \left(\Delta \varepsilon_r'+\varepsilon_r'\right)\right.\\ \nonumber
  & \left.-2 \zeta n_{\rm sp}  (\partial_z\Delta\varepsilon_r') \sqrt{-g_{00}\left(r_\oplus+z\right)} (\Delta \varepsilon_r+\varepsilon_r)+\zeta n_{\rm sp} (\partial_z\Delta\varepsilon_r) \sqrt{-g_{00}\left(r_\oplus+z\right)} \left(\Delta \varepsilon_r'+\varepsilon_r'\right)\right.\\ \nonumber
    &\left.-2n_{\rm sp} (\partial_z\zeta) \sqrt{-g_{00}\left(r_\oplus+z\right)} (\Delta \varepsilon_r+\varepsilon_r) \left(\Delta \varepsilon_r'+\varepsilon_r'\right)-2 \zeta n_{\rm sp} \partial_z\sqrt{-g_{00}(r_\oplus+z)} (\Delta \varepsilon_r+\varepsilon_r) \left(\Delta \varepsilon_r'+\varepsilon_r'\right)\right],
\end{align}
\begin{align}
    {\partial^2_z\kt_0\approx}&{\kappa^{\rm b}_0 \sqrt{-g_{00}(r_\oplus)} \left(\partial_z^2 n_{\rm sp} \sqrt{\frac{\Delta\varepsilon_r}{\varepsilon _r}+1}+\frac{ \partial_z n_{\rm sp} (\partial_z\Delta\varepsilon_r)}{\varepsilon _r \sqrt{\frac{\Delta\varepsilon_r}{\varepsilon _r}+1}}+n_{\rm sp} \left(\frac{\partial_z^2\Delta \varepsilon_r}{2 \varepsilon _r \sqrt{\frac{\Delta\varepsilon_r}{\varepsilon _r}+1}}-\frac{(\partial_z\Delta\varepsilon_r)^2}{4 \varepsilon _r^2 \left(\frac{\Delta\varepsilon_r}{\varepsilon _r}+1\right){}^{3/2}}\right)\right)}\\ \nonumber
    &-\frac{\bar{\omega}_0^2 \sqrt{-g_{00}(r_\oplus+z)} \zeta (\partial_z^2n_{\rm sp}) \left(\Delta\varepsilon_r'+\varepsilon_r'\right)}{2 c \sqrt{\Delta \varepsilon_r+\varepsilon_r}}-\frac{\bar{\omega}_0^2 (\partial_zn_{\rm sp}) \left(2 \zeta (\partial_z\sqrt{-g_{00}(r_\oplus+z)}) (\Delta \varepsilon_r+\varepsilon_r) \left(\Delta\varepsilon_r'+\varepsilon_r'\right)-\sqrt{-g_{00}(r_\oplus+z)} \zeta \partial_z\Delta\varepsilon_r \left(\Delta\varepsilon_r'+\varepsilon_r'\right)\right)}{2 c (\Delta \varepsilon_r+\varepsilon_r)^{3/2}}\\ \nonumber
    &-\frac{\bar{\omega}_0^2 (\partial_zn_{\rm sp}) \left(2 \sqrt{-g_{00}(r_\oplus+z)} \zeta (\partial_z\Delta\varepsilon_r') (\Delta \varepsilon_r+\varepsilon_r)+2 \sqrt{-g_{00}(r_\oplus+z)} (\partial_z\zeta) (\Delta \varepsilon_r+\varepsilon_r) \left(\Delta\varepsilon_r'+\varepsilon_r'\right)\right)}{2 c (\Delta \varepsilon_r+\varepsilon_r)^{3/2}}\\ \nonumber
    &-\frac{\bar{\omega}_0^2 n_{\rm sp} \left(4 (\Delta \varepsilon_r+\varepsilon_r) \left(2 (\partial_z\sqrt{-g_{00}(r_\oplus+z)}) (\Delta \varepsilon_r+\varepsilon_r)-\sqrt{-g_{00}(r_\oplus+z)} \partial_z\Delta\varepsilon_r\right) \left(\zeta (\partial_z\Delta\varepsilon_r')+(\partial_z\zeta) \left(\Delta\varepsilon_r'+\varepsilon_r'\right)\right)\right)}{8 c (\Delta \varepsilon_r+\varepsilon_r)^{5/2}}\\ \nonumber
    &-\frac{\bar{\omega}_0^2 n_{\rm sp} \left(-\zeta \left(\Delta\varepsilon_r'+\varepsilon_r'\right) \left(-4 (\partial_z^2\sqrt{-g_{00}(r_\oplus+z)}) (\Delta \varepsilon_r+\varepsilon_r)^2+4 (\partial_z\sqrt{-g_{00}(r_\oplus+z)}) \partial_z\Delta \varepsilon_r (\Delta \varepsilon_r+\varepsilon_r)+2 \sqrt{-g_{00}(r_\oplus+z)} \partial_z^2\Delta \varepsilon_r (\Delta \varepsilon_r+\varepsilon_r)\right)\right)}{8 c (\Delta \varepsilon_r+\varepsilon_r)^{5/2}}\\ \nonumber
    &-\frac{\bar{\omega}_0^2 n_{\rm sp} \left(-\zeta \left(\Delta\varepsilon_r'+\varepsilon_r'\right) \left(-3 \sqrt{-g_{00}(r_\oplus+z)} (\partial_z\Delta\varepsilon_r)^2\right)\right)}{8 c (\Delta \varepsilon_r+\varepsilon_r)^{5/2}}-\frac{\bar{\omega}_0^2 n_{\rm sp} \left(4 \sqrt{-g_{00}(r_\oplus+z)} (\Delta \varepsilon_r+\varepsilon_r)^2 \left(\zeta (\partial_z^2\Delta\varepsilon_r')+2 (\partial_z\Delta\varepsilon_r') (\partial_z\zeta)+(\partial_z^2\zeta) \left(\Delta\varepsilon_r'+\varepsilon_r'\right)\right)\right)}{8 c (\Delta \varepsilon_r+\varepsilon_r)^{5/2}}\\ \nonumber
\end{align}
}

Considering that 
\begin{equation}\label{fullapprox2}
    \Delta \varepsilon_r=-\frac{\varepsilon_r^2\Delta (\varepsilon_r^{-1})}{1+\varepsilon_r\Delta (\varepsilon_r^{-1})},
\end{equation}
with
\begin{align}\label{deltaminus}
    \Delta (\varepsilon_r^{-1})=\frac{c^2 P_{1122} r_S \left(\frac{z}{\left(1-\frac{r_S}{4 (L+r_\oplus)}\right) \left(\frac{r_S}{4 (L+r_\oplus)}+1\right)^3}-\frac{(z-L)^2-L^2}{(L+r_\oplus) \left(\frac{r_S}{4 (L+r_\oplus)}+1\right)^6}\right)}{2 c_s^2 (L+r_\oplus)^2}
\end{align}
we also have
\begin{align}
    &\partial_z \Delta \varepsilon_r=-\frac{\varepsilon_r^2 \partial_z\Delta(\varepsilon_r^{-1})}{(\varepsilon_r \Delta(\varepsilon_r^{-1})+1)^2}\\
    &\partial^2_z \Delta \varepsilon_r=\frac{\varepsilon_r^2 \left(2 \varepsilon_r (\partial_z\Delta(\varepsilon_r^{-1}))^2-(\varepsilon_r \Delta(\varepsilon_r^{-1})+1) \partial_z^2\Delta(\varepsilon_r^{-1})\right)}{(\varepsilon_r \Delta(\varepsilon_r^{-1})+1)^3}.
\end{align}

Moving on, for $\kappa_1$ we need 
\begin{align}
    &\kappa_1(\omega_0)=(n(\omega_0)+\omega_0\partial_\nu n(\nu)|_{\omega_0})/{c}
\end{align}
We proceed with the same approximation at first order in $\zeta$ as done above. We get
\begin{align}
    \kappa_1(\omega_0)\approx & {{c^{-1}\left(\sqrt{\varepsilon_r+\Delta \varepsilon_r}+\bar{\omega}_0\frac{\epsilon_r'+\Delta\varepsilon_r'}{2\sqrt{\varepsilon_r+\Delta \varepsilon_r}}{\frac{\sqrt{-g_{00}(r_{\oplus})}}{\sqrt{-g_{00}(r_{\oplus}+z)}}}\right)}}\\ \nonumber 
    & +c^{-1}\left[\zeta \left(\frac{1}{2} \bar{\omega}_0 \left(\frac{-\bar{\omega}_0 \Delta \varepsilon_r''-\bar{\omega}_0 \varepsilon_r''}{\sqrt{\Delta \varepsilon_r+\varepsilon_r}}+\left(\frac{\bar{\omega}_0 \Delta \varepsilon_r'+\bar{\omega}_0 \varepsilon_r'}{2 (\Delta \varepsilon_r+\varepsilon_r)^{3/2}}-\frac{1}{\sqrt{\Delta \varepsilon_r+\varepsilon_r}}\right) \left(\Delta \varepsilon_r'+\varepsilon_r'\right)\right)\right)\right],\\ \nonumber
\end{align}
where all quantities on the right-hand side are evaluated at $\bar{\omega}_0$. 

{Following the same notation as before, we indicate with $\kappa_1^{\rm b}$ the tabulated optical parameter for the material without photoelasticity. This tabulated quantity enters the previous expression through $\varepsilon_r'=\chi_1'(\bar{\omega}_0)$. Indeed, from $\kappa^{\rm b}_1=(n_0+\bar{\omega}_0\partial_\nu n_0(\nu)|_{\bar{\omega}_0})/{c}$, we have
}
\begin{equation}
    \kappa^{\rm b}_1=c^{-1}(n_0+\bar{\omega}_0\partial_\nu n_0(\nu)|_{\bar{\omega}_0})={c^{-1}}\left(n_0+\bar{\omega}_0\frac{\chi_1'}{2 n_0}\right), 
\end{equation}
from which we can read $\chi_1'= {2c^2}\left(-(\kappa_0^{\rm b})^2+\kappa^{\rm b}_0\kappa^{\rm b}_1\bar{\omega}_0\right)/{\bar{\omega}_0^3}$.

We thus remain with identifying $\Delta\varepsilon_r'$. Let us consider the full form of $\Delta\varepsilon_r$ in {eq.~\eqref{fullapprox2}
and notice that} $\Delta (\varepsilon_r^{-1})$ {in there, as given in eq.~\eqref{deltaminus},} does not depend on the frequency but only on the stresses and strains. Thus we get,
\begin{equation}
    \Delta \varepsilon_r'=-\frac{\varepsilon_r'\varepsilon_r\Delta (\varepsilon_r^{-1})\left(2+\varepsilon_r\Delta (\varepsilon_r^{-1})\right)}{(1+\varepsilon_r\Delta (\varepsilon_r^{-1}))^2},
\end{equation}
with \begin{equation}
    \varepsilon_r'=\chi_1',
\end{equation}

Finally, under the same assumption as before, we have
\begin{align}
    &\kappa_2(\omega_0)={c^{-1}}(2 n(\nu)'|_{{\omega}_0}+{\omega}_0 n(\nu)''|_{{\omega}_0}).
\end{align}
Thus we end up with 
\begin{align}
    \kappa_2(\omega_0)\approx &{{c^{-1}\left(\frac{\Delta \varepsilon _r'+\varepsilon _r'}{\sqrt{\Delta \varepsilon _r+\varepsilon _r}}+\bar{\omega}_0\frac{2 \left(\Delta \varepsilon _r+\varepsilon _r\right) \left(\Delta \varepsilon _r''+\varepsilon _r''\right)-\left(\Delta \varepsilon _r'+\varepsilon _r'\right){}^2}{4 \left(\Delta \varepsilon _r+\varepsilon _r\right){}^{3/2}}{\frac{\sqrt{-g_{00}(r_{\oplus})}}{\sqrt{-g_{00}(r_{\oplus}+z)}}}\right)}}\\ \nonumber
    &+\frac{\zeta \bar{\omega}_0 \left(-\left(-3 \bar{\omega}_0 \left(\Delta \varepsilon_r'+\varepsilon_r'\right)+4 \Delta \varepsilon_r+4 \varepsilon_r\right) \left(2 (\Delta \varepsilon_r+\varepsilon_r) \left(\Delta \varepsilon_r''+\varepsilon_r''\right)-\left(\Delta \varepsilon_r'+\varepsilon_r'\right)^2\right)\right)}{8 c (\Delta \varepsilon_r+\varepsilon_r)^{5/2}}\\ \nonumber
    &-\frac{\zeta \bar{\omega}_0^2 \left(\Delta \varepsilon_r''' +\varepsilon_r'''\right)}{2 c (\Delta \varepsilon_r+\varepsilon_r)^{1/2}},
\end{align}
where all the quantities on the right-hand side are evaluated at $\bar{\omega}_0$.

As before, {indicating the tabulated optical property of the material with $\kappa^{\rm b}_2$ and $\kappa^{\rm b}_3$ without photoelasticity, it is immediate to derive an expression for $ \chi_1''(\bar{\omega}_0)$}
\begin{equation}
    \chi_1'' = \frac{2c^2}{\bar{\omega}_0^4}\left(3(\kappa^{\rm b}_0)^2-4\kappa^{\rm b}_0 \kappa^{\rm b}_1 \bar{\omega}_0 + \left(\kappa^{\rm b}_1\bar{\omega}_0\right)^2 +\kappa^{\rm b}_0\kappa^{\rm b}_2 \bar{\omega}_0^2\right),
\end{equation}
and 
\begin{align}\label{chi3}
    \chi_1'''=\frac{2 c^2 \left(-12 (\kappa^{\rm b}_0)^2-6 \bar{\omega}_0^2 \left(\kappa^{\rm b}_0 \kappa^{\rm b}_2+(\kappa^{\rm b}_1)^2\right)+\bar{\omega}_0^3 (\kappa^{\rm b}_0 \kappa^{\rm b}_3+3 \kappa^{\rm b}_1 \kappa^{\rm b}_2)+18 \kappa^{\rm b}_0 \kappa^{\rm b}_1 \bar{\omega}_0\right)}{\bar{\omega}_0^5}
\end{align}
while
\begin{equation}
    \Delta \varepsilon_r''=\frac{\Delta (\varepsilon_r^{-1}) \left(-\varepsilon _r \left(\Delta (\varepsilon_r^{-1}) \varepsilon _r+1\right) \left(\Delta (\varepsilon_r^{-1}) \varepsilon _r+2\right) \varepsilon _r''-2 \varepsilon _r'{}^2\right)}{\left(\Delta (\varepsilon_r^{-1}) \varepsilon _r+1\right){}^3},
\end{equation}
and
\begin{align}
    & \Delta \varepsilon_r'''= -\frac{\Delta (\varepsilon_r^{-1})  \left(\varepsilon_r \varepsilon_r''' (\Delta (\varepsilon_r^{-1})  \varepsilon_r+1)^2 (\Delta (\varepsilon_r^{-1})  \varepsilon_r+2)-6 \Delta (\varepsilon_r^{-1})  \varepsilon_r'^3+6 (\Delta (\varepsilon_r^{-1})  \varepsilon_r+1) \varepsilon_r' \varepsilon_r''\right)}{(\Delta (\varepsilon_r^{-1})  \varepsilon_r+1)^4},
\end{align}
with \begin{equation}
    \varepsilon_r''=\chi_1''\,\,\text{and}\,\, \varepsilon_r'''=\chi_1'''.
\end{equation}
In Eq.\eqref{chi3}, we also neglect $\kappa^{\rm b}_3$ since this term is negligible.
}

{
\section{Width of the pulse}
While until now we have considered only the effect of a gravitational field on the propagation velocity of the optical pulse, we can also look at the width of the pulse while it propagates. In the horizontal case, the width remains constant, as can be seen from the analytical solution of eq.~\eqref{eq:soliton_diffeq_hor1}. In the vertical propagation case, however, this is no longer true. From Fig.~\ref{fig:width}, we see that spacetime effects, in conjunction with photoelasticity, reduce the width of the pulse. This is clearly negligible for realistic values of $r_S$, and it becomes relevant only at extreme values but shows, nonetheless, that gravity has a focusing effect on the propagating pulse.   
\begin{figure}
    \centering
\includegraphics[scale=0.5]{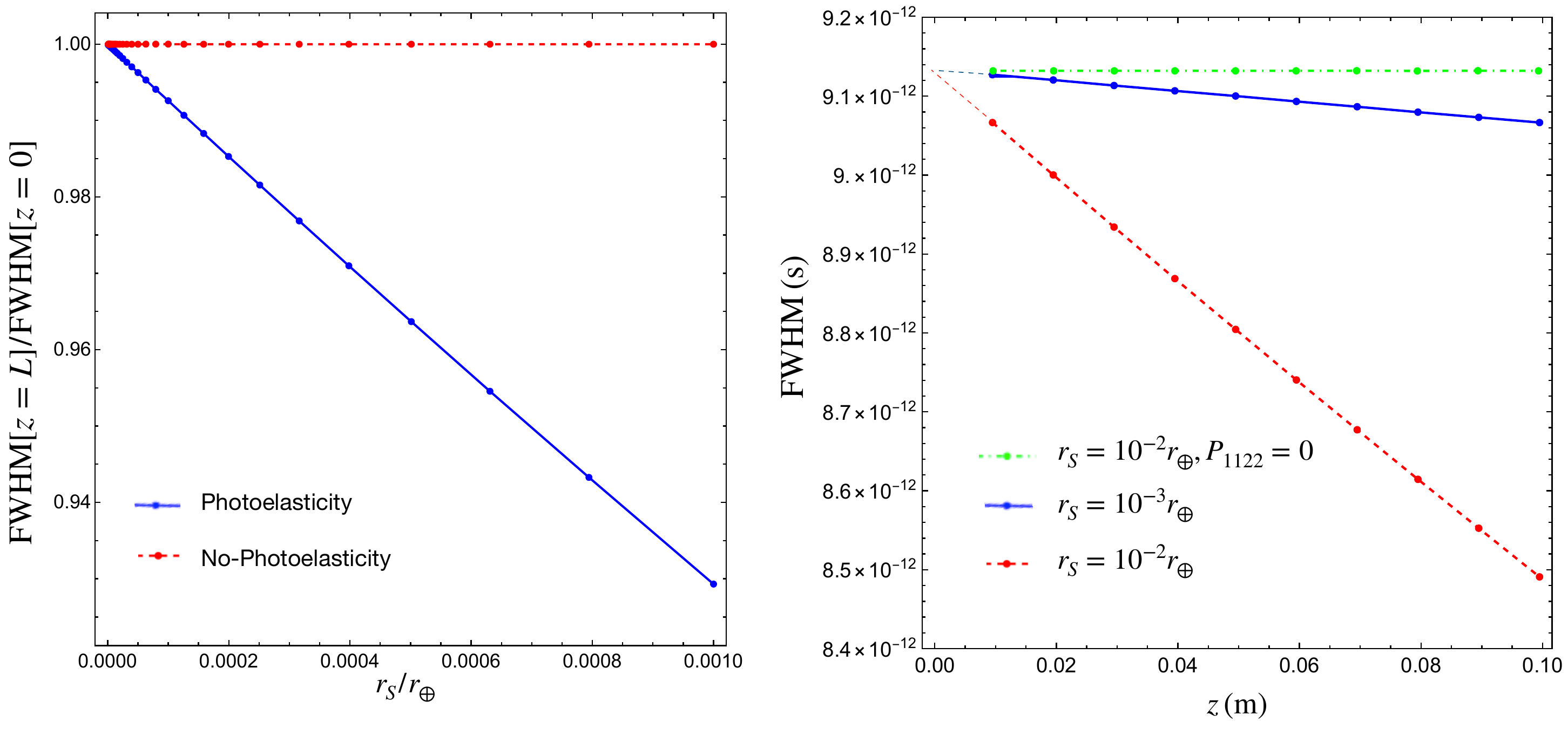}
    \caption{\textbf{Left:} {FWHM (full width at half maximum) of the pulse} at $z=L$ normalized by the {FWHM of the initial pulse at $z=0$} as a function of $r_S/r_\oplus$ for an initial pulse with $T_0=40 \cdot 10^{-13}$\,s propagating for 0.1\,m. {The blue solid line shows the case including the effect of photoelasticity, the red dotted line represents the case without photoelasticity.} \textbf{Right:} pulse FWHM (in seconds) as a function of the propagation distance $z$ for two different values of $r_S$, again using a pulse with $T_0=40 \cdot 10^{-13}$\,s, longer than the one previously considered, for better numerical precision. The green dot-dashed line represents the case without photoelasticity.}
    \label{fig:width}
\end{figure}
}

\end{document}